\documentclass[manuscript]{aastex62}

\usepackage{rotating}
\usepackage{graphicx}

\newcommand{\RSUN}{R$_\sun$}

\submitjournal{Astronomical Journal}
\accepted{October 23, 2023}

\shorttitle{Plasma Tail of Comet S3}
\shortauthors{Li et al.}

\begin{document}

\title{The Wagging Plasma Tail of Comet C/2020 S3 (Erasmus)}

\correspondingauthor{Jing Li}
\email{jingli@epss.ucla.edu}

\author[0000-0002-0982-7309]{Jing Li}
\affil{Department of Earth, Planetary and Space Sciences, University of California, Los Angeles, CA 90095, USA}
\author[0000-0002-4676-2196]{Yoonyoung Kim}
\affiliation{Department of Earth, Planetary and Space Sciences, University of California, Los Angeles, CA 90095, USA}
\author[0000-0003-0951-7762]{David Jewitt}
\affiliation{Department of Earth, Planetary and Space Sciences, University of California, Los Angeles, CA 90095, USA}
%


\begin{abstract}
Long-period comet C/2020 S3 (Erasmus) reached perihelion at 0.398 au on UT 2020 December 12.67, making it a bright, near-Sun object. Images taken between mid-November and December 2020 using the HI-1 camera and COR2 coronagraph onboard STEREO-A, as well as the LASCO/C3 coronagraph onboard SoHO, show significant variations in the plasma tail position angles. To analyze these variations, a simple technique was developed to calculate the aberration angles. These angles are defined as the angle between the sun-comet line and the tail axis, measured in the orbital plane. The aberration angles were found to range from $1.2^\circ$ to $46.8^\circ$, with an average (median) value of approximately $20.3^\circ$ ($16.3^\circ$). By considering the aberration angles, the solar wind radial velocities during the observations were inferred to range from 73.9 km/s to 573.5 km/s, with mean (median) values of approximately 205.5 km/s (182.3 km/s). Throughout the observations, two periods were identified where the tails showed forward tilting, which cannot be explained by aberration alone. In one case, this anomalous position angle was sustained for at least 11 days and is possibly due to co-rotating interaction regions. In the other case, the tail exhibited dramatic excursions from 180$^\circ$ to 150$^\circ$ back to 210$^\circ$ over a limited period of around 34 hours. This behavior is tentatively explained as a consequence of the interaction with a halo Coronal Mass Ejection that was launched from NOAA 12786 and arrived at comet C/2020 S3 during the time when the tail displayed its wagging behavior.
\end{abstract}



\section{Introduction} \label{sec:intro}

Long-period comet C/2020 S3 (Erasmus) (hereafter ``S3'') was discovered on UT 2020 September 17 by the Asteroid Terrestrial-impact Last Alert System (ATLAS) sky survey\footnote{\url{https://minorPlanetcenter.net/mpec/K20/K20SB9.html}}. It has osculating semimajor axis $a$ = 194.9 au, eccentricity $e$ = 0.998 and inclination $i$ = 19.9$\degr$. (The barycentric elements are very similar at $a$ = 190 au, $e$ = 0.998 and $i$ = 19.9\degr). As such, S3 has a likely origin in the Oort cloud but it is not a dynamically new object (for which $a > 10^4$ au is required). Comet S3 reached its perihelion on  UT 2020 December 12.67 at 0.398 au. The perihelion water production rate  was $Q_{H_2O} = 4\times10^{29}$ s$^{-1}$ \citep{combi23}, corresponding to a mass production rate in gas $\sim$12,000 kg s$^{-1}$.  As the comet approached its perihelion, the solar elongation as seen from Earth decreased from $46\degr$ (mid-November) down to $10\degr$ (mid-December) rendering ground-based observation increasingly difficult. However, the comet was captured by the Sun-observing STEREO-A and SoHO spacecraft both before and after its perihelion, providing the dataset we analyze here. \\

As is well known from the study of other comets, dust and gas move differently once released from the nucleus. Cometary dust is expelled from the sublimating nucleus by gas drag, typically attaining a size-dependent terminal speed small compared to the thermal speed $V_{th}$ of gas molecules.  Once ejected, the motion of dust is controlled  by solar gravity and radiation pressure. The force due to solar wind drag is at least an order of magnitude smaller than radiation pressure, and can be ignored \citep{2003ApJ...594.1049R}. Comet dust tails are usefully represented by syndynes marking the locus of positions of particles of a given size released continuously from the comet and synchrones which show the locus of positions of particles having a range of sizes but released at a common time  \citep{Finson68}.  \\

Different forces influence the ion tails. Gas molecules become ionized primarily by solar ultraviolet photons and are swept up by the passing solar wind to form the comet's plasma tail, as first explained by \citet{1951ZA.....29..274B}. In brief, gas molecules leave the comet nucleus with a speed comparable to the thermal speed appropriate to the temperature of the sublimating surface, typically $V_{th} \sim1$ km s$^{-1}$ at 1 au, while the orbital motion of the comet is more comparable to the Kepler speed, $v_K \sim $ a few tens of km s$^{-1}$ at 1 au.    Both speeds are small compared to that of the solar wind, which sweeps past the comet at a range $\sim200$ to $\sim 750$ km s$^{-1}$ \citep{1968SSRv....8..690H}, and expands outward from the Sun in a roughly radial direction.  In  interplanetary space, the plasma pressure carried by the solar wind is bigger than the magnetic field pressure so that the magnetic fields from the Sun are frozen-in to and carried by the solar wind. As a result,  cometary ions are picked up by interplanetary magnetic fields (IMF) via the Lorentz force and form a plasma tail that is characteristically aligned usually close to the sun-comet line.\\

For these reasons, the dust and plasma tails are typically spatially separated and appear distinct in the plane of the sky.  Whereas the plasma tails are straight and nearly radial to the Sun, dust tails are commonly diffuse and curved in the plane of the sky as a result of the combined radiation pressure and orbital motion. The directions of both types of tail should change smoothly in response to the changing observer-sun-comet geometry.  A third, very narrow and typically faint type of tail has been associated with resonance fluorescence from neutral sodium atoms \citep{2002AdSpR..29.1187C}.  Sodium atoms have very large solar radiation pressure acceleration (e.g.~100s of times solar gravity), and so closely track the antisolar direction. For example, a sodium tail was recently reported  on asteroid 3200 Phaethon near  perihelion using the same instruments as we use in this work \citep{2023AJ....165...94H,2023PSJ.....4...70Z}.  However, our data provide no clear evidence for a neutral sodium tail perhaps for the reason that the S3 is not as bright as comet C/1995 O1 Hale-Bopp \citep{1997ApJ...490L.199C}. \\

Although plasma tails are roughly aligned with sun-comet lines, they often show  dynamic fine structures imprinted by the local solar wind environment. In fact, the existence of the solar wind was first realized from the appearance of cometary plasma tails \citep{1951ZA.....29..274B,1957Tell....9...92A,1958ApJ...128..664P,1968ARA&A...6..267B}. It is frequently observed that plasma tails appear cut-off in so-called Disconnection Events (DEs) \citep{1978ApJ...223..655N,1985AdSpR...5l..65O,2000A&A...358..759W,2005EM&P...97..399V}. Suggested causes of DEs include variable ion production rates in the cometary gas \citep{1976Icar...29..147I}, interplanetary shock waves, the comet passing through a heliospheric current sheet (i.e.~a sector boundary \citep{1979ApJ...234..723N}), and the impact of coronal mass ejections \citep{2007ApJ...668L..79V,2009ApJ...696L..56J}. For their visibly disturbed plasma tails, sungrazing comets come close to the Sun within only a few tens of solar radii. They are studied as probes to infer solar coronal physical conditions    \citep{2001ApJ...558..403U,2012ApJ...760...18B,2013Sci...340.1196D,2018SSRv..214...20J,2020ApJ...897...87C,2022ApJ...938...20N,2022JGRA..12729799R}. These comets also provide good opportunities to study the constituents of the cometary coma and plasma tails, for example, with the UltraViolet Coronagraph Spectrometer (UVCS) \citep{1995SoPh..162..313K} onboard the SoHO satellite \citep{2003Sci...302.1949P,2018ApJ...858...19R,2022ApJ...926...93R}. These observations show that the cometary emission lines, H I Ly$\alpha$ is the strongest UV emission line in the cometary coma due to sunlight scattered by H atoms produced from  H$_2$O photodissociation. An ion tail consisting of C III has also been observed by UVCS as covered in the works by \citet{2003Sci...302.1949P,2018ApJ...858...19R,2022ApJ...926...93R}.\\

Because the plasma tail appearances reflect the local solar wind environment, a direct application of the comet plasma tails is to infer the solar wind radial velocity using the sky-plane position angles of the tails. Because of the comet motion, the tail forms an angle with respect to the sun-comet line, defining the so-called aberration angle, $\varepsilon$. In the comet orbital plane, the angle is simply written as \citep{1957Tell....9...92A} in the present of solar wind,
\begin{equation}
\varepsilon = tan^{-1}\left(\frac{v_t}{w_r-v_r}\right)
\label{eq:alfven}
\end{equation}

\noindent where $v_t$ and $v_r$ represent the perpendicular and radial components of the comet orbital velocities; $w_r$ represents the solar wind radial velocity.  Past measurements of 60 comets yield typical aberration angles $\varepsilon \sim5\degr$ \citep{1966ApJS...13..125B,1968ARA&A...6..267B}. However, recent measurements show that $\varepsilon$ can span a wider range, $5\degr \lesssim \varepsilon \lesssim30\degr$ \citep{2020ApJ...897...87C}. \\

The aberration angle can be obtained from the comet tail in various ways based on the fact that the plasma tail orientation is dictated by the comet orbital motion and the solar wind. One way is to fit to the tail position angles by assuming solar wind velocity at the location of the comets. This is achieved by a coordinate transfer from the orbital plane to the observer's projected sky plane \citep{1966ApJS...13..125B,1978ApJ...221.1014N}. Several comets have been observed by Solar Mass Ejection Imager (SMEI) on the Coriolis spacecraft, which follows a polar orbit above the Earth's terminator. The combined fisheye sky map shows the entire sky, revealing comet tails as long as 0.5 au. The aberration angles along with solar wind speeds are obtained at multiple locations along the tail in a coordinate system originating at the comets \citep{2008ApJ...677..798B,2010ApJ...713..394C,2013AIPC.1539..364J}. Using STEREO twin spacecraft, \citet{2020ApJ...897...87C} reconstruct the comet images in three dimensional space, and obtain the comet aberration angles and solar wind velocities. They adapted their method to use  single telescope measurements to determine the aberration angles \citep{2022ApJ...928..121C}. A different technique using time-series images of the comet is employed by \citet{2022JGRA..12729799R} to infer the solar wind radial velocity from the comet aberration angles. In this paper, we report near-Sun observations of S3,  focusing on the plasma tail. We estimate the comet aberration angles and solar wind speeds from the position angles and the known geometry of observation. We will introduce a new method to determine the aberration angles. \\

The paper is organized as follows: In section \ref{sec:obs}, the STEREO-A and SoHO data are described. In section \ref{sec:tails}, measurements of the comet tails are described and analyzed. The results are discussed in section \ref{sec:discussion}, and a summary is given in section \ref{sec:summary}.\\

\section{Observations} \label{sec:obs}
Comet S3 was observed using two spacecraft: the Solar Terrestrial Relations Observatory-A (STEREO-A) \citep{2008SSRv..136....5K}, and the Solar and Heliospheric Observatory (SoHO) \citep{1995SoPh..162....1D}.  In chronological order, the comet was first recorded by a wide field camera (HI-1), then by a narrow field coronagraph (COR2), both onboard STEREO-A \citep{2008SSRv..136...67H,2009SoPh..254..387E}, then by a coronagraph (C3) which is a part of the Large Angle and Spectrometric Coronagraph Experiment (LASCO)  \citep{1995SoPh..162..357B} onboard SoHO.  The observational times and instrument parameters are listed in Table (\ref{tbl:cameras}). The two STEREO-A cameras have similar spectral coverage with filter bandwidths from 630 to 730 nm for HI-1 and from 650 to 750 nm for COR2, but they have different fields  of view and angular resolutions.  The LASCO/C3 camera has a filter wheel containing five filters: a wide bandpass filter (400 to 835 nm), blue, orange, deep red, and infrared filters. The clear filter is normally used for the observations, which is the case for the current study. \\

\begin{figure}
\centering
\includegraphics[width=\textwidth]{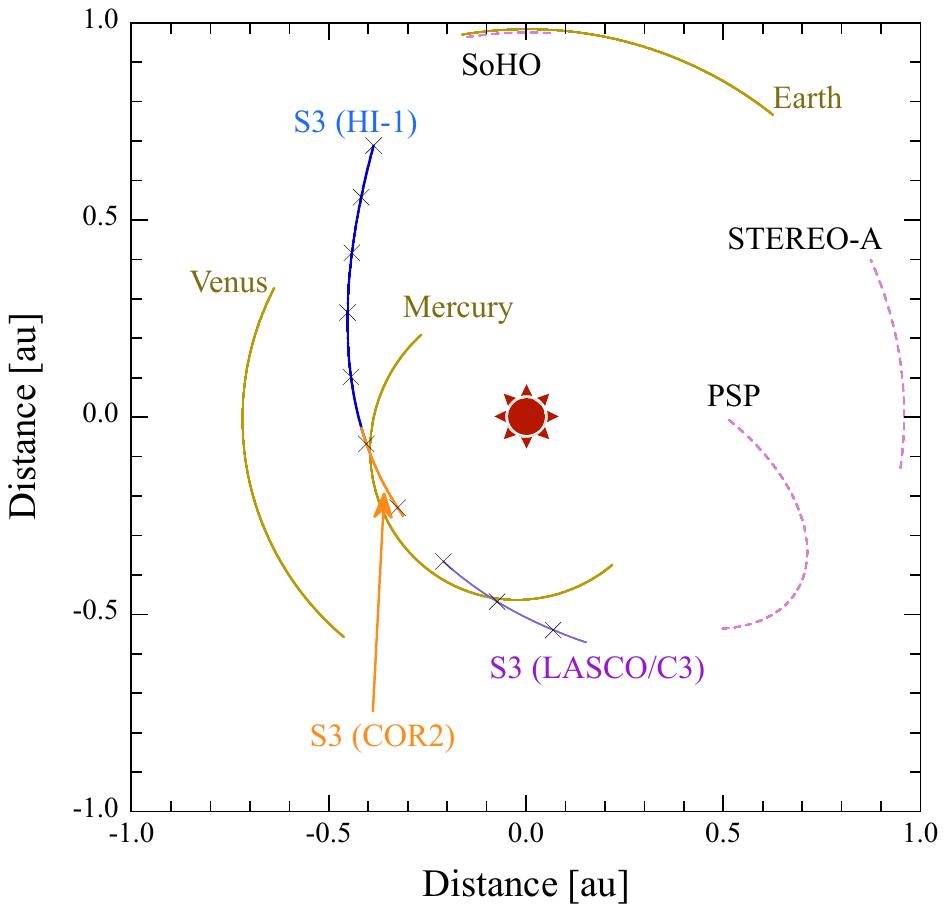}
\caption{Plan view of the orbital paths of comet S3, the Planets, STEREO-A, SoHO and Parker Solar Probe (PSP) in the heliocentric cartesian coordinate during the observing period, from mid-November to end-December 2020. All objects orbit  counterclockwise. The comet trajectory is plotted as blue, orange and purple arcs corresponding to the observations  with HI-1, COR2 and LASCO/C3 cameras, respectively. Tick marks ``X'' are plotted every 5 days on the S3 path starting with DOY 318.  Three terrestrial planets are represented by brown color. The spacecraft are plotted with dotted pink curves.  \label{fig:orbits}}
\end{figure}

\begin{figure}
\includegraphics[width=\textwidth]{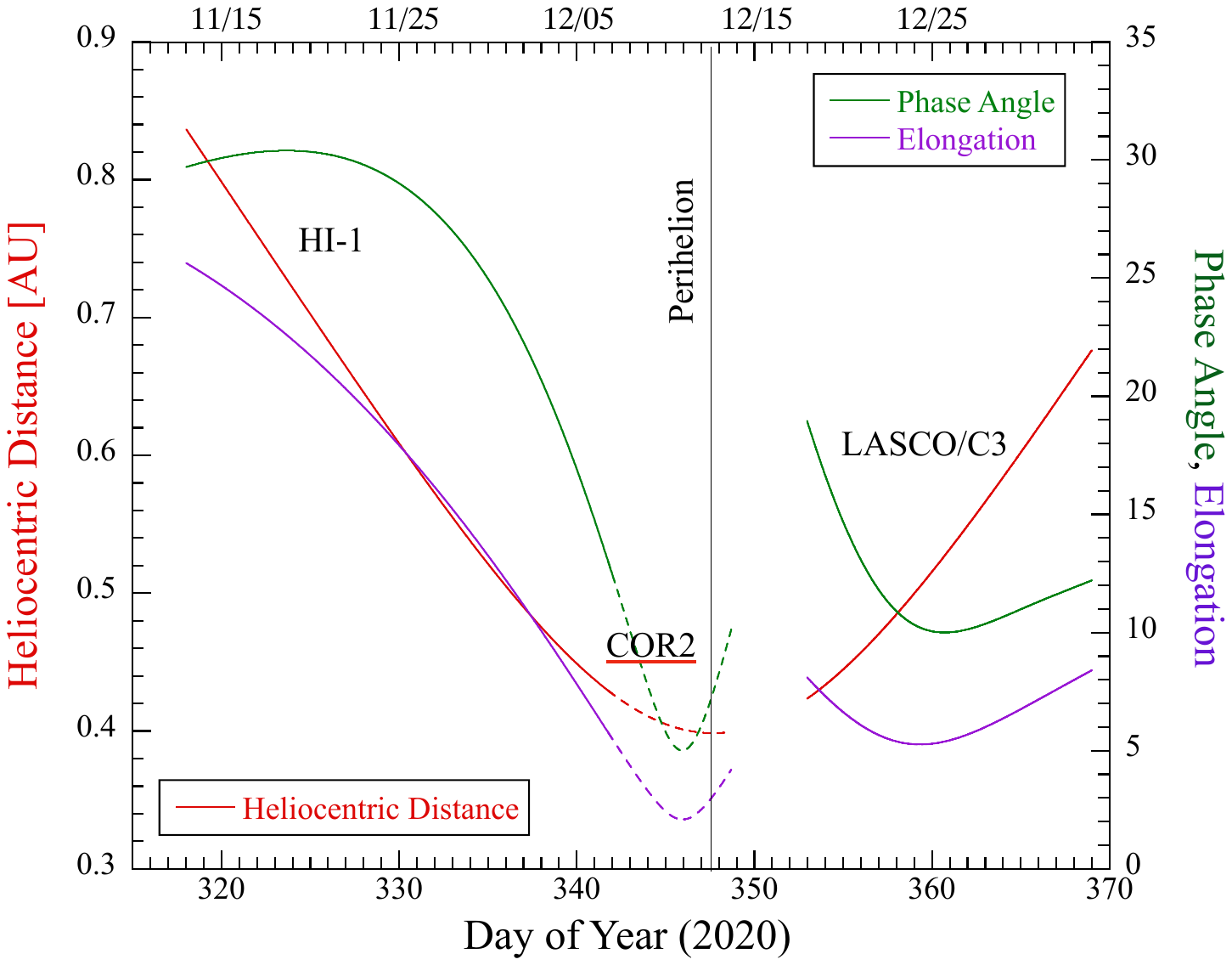}
\caption{Observational geometry as a function of time. The S3 observations are made with HI-1 wide field camera (solid lines) and COR2 coronagraph (dotted lines) on STEREO-A. After DOY 350, S3 is detected by LASCO/C3 coronagraph on SoHO (solid lines). Perihelion is at UT 2020 December 12.67, indicated by a vertical line. Heliocentric distance, phase angle and elongation are represented by red, green and purple colors, respectively. Calendar dates are indicated along the top axis. \label{fig:e-hi1-cor2}}
\end{figure}

The comet observations lasted from 2020 mid-November to 2021 January. STEREO-A is a sun-orbiting spacecraft drifting far from Earth while SoHO orbits the Earth-Sun L1 Lagrangian point; as a result, the observing geometry at any time differs between the two.  For easy reference, Figure (\ref{fig:orbits}) shows a plan view of  the orbital path of S3, together with the terrestrial planets and the  STEREO-A, SoHO and, for reference, the Parker Solar Probe (PSP) spacecraft  during the observing period. Figure (\ref{fig:e-hi1-cor2}) shows the temporal variation of the  geometric parameters as S3 crossed HI-1 (solid curves), COR2 (dotted curves) and C3 (solid curves at later time) cameras. S3 exited the COR2 field-of-view almost exactly one day after perihelion. \\

The comet stayed in the wide angle HI-1 camera field for 24 days from UT 2020 November 13.0 to December 7.0.  It then crossed the narrower field of view of COR2  from UT December 6.5 to 13.7. Starting about 4.3 days later, it was observed by LASCO/C3, from December 18.0 to 2021 January 3.0. The comet image quality becomes too poor to obtain reliable measurements of the tail position angles after  2021 January 1. Therefore, we use only the combined observational data between 2020 mid-November and the end of December.  S3 was not observed at the same time from different vantage points and, therefore, a stereoscopic view  like that obtained by \citet{2020ApJ...897...87C} with the twin STEREO spacecraft is not available for this work. \\

\begin{figure}
\centering
\includegraphics[width=0.95\textwidth]{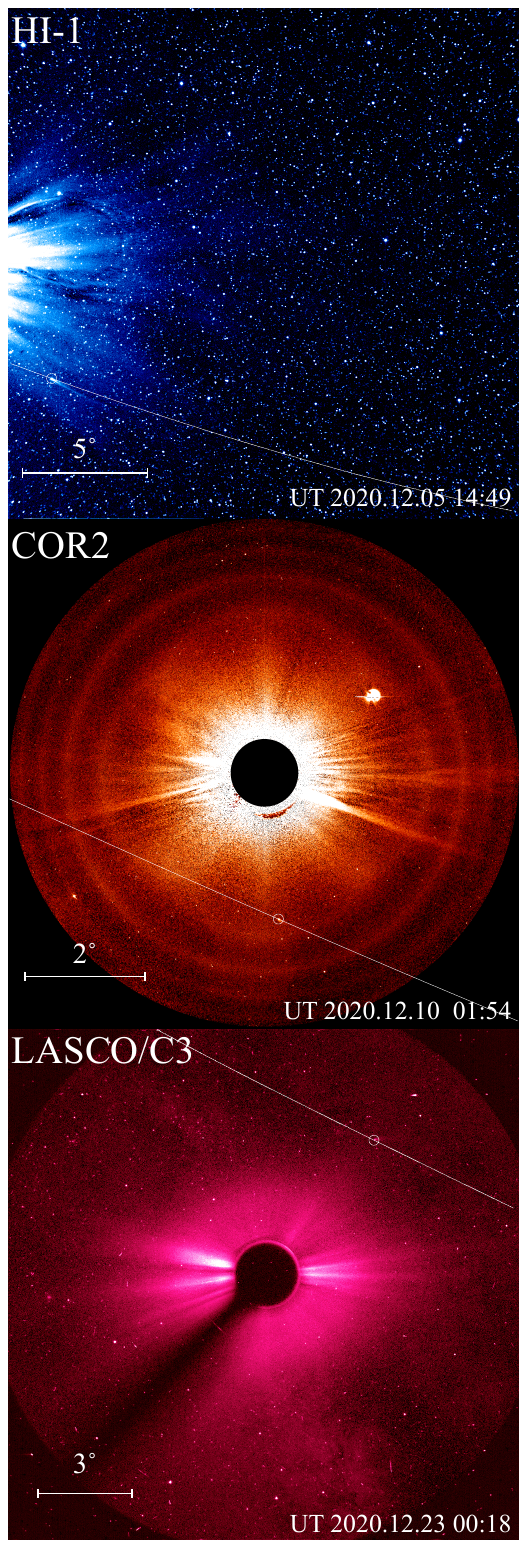}
\caption{Full coronal images from the wide-field camera HI-1 (top), the narrow-field camera COR2 (middle) both on STEREO-A, and the LASCO/C3 coronagraph (bottom) on SoHO. The projected comet paths are plotted with solid white curves. The comet positions at the times of images are circled on each image. \label{fig:hi1-cor2-c3}}
\end{figure}

We downloaded level-0 data from the STEREO-A site\footnote{\url{https://stereo-ssc.nascom.nasa.gov/data/ins\_data/secchi/L0/a/img/}} and further processed them to level-1 using the standard IDL procedure, {\it secchi\_prep}, in the SolarSoftWare (SSW) package \citep{1998SoPh..182..497F}.  This step included  flat-field and stray light corrections. We applied a  data reduction technique similar to that described in \citet{2010AJ....140.1519J}, with the purpose being to obtain ``clean'' images free of stars and transient features such as coronal mass ejections. This was achieved by subtracting a rolling median image of the corona from individual images within the sequence. The comet was located in the STEREO-A images using the World Coordinate System (WCS) provided in the STEREO-A FITS headers \citep{2010SoPh..261..215T} combined with JPL Horizons comet ephemerides in celestial coordinates.  The examples of final processed images from HI-1 and COR2 cameras are shown in Figure (\ref{fig:hi1-cor2-c3}) in the top and middle panels. \\

We downloaded the C3/LASCO data from the LASCO level\_0.5 data site \footnote{\url{https://umbra.nascom.nasa.gov/pub/lasco\_level05/960430/c3/}}. To remove the background coronal emission, a minimal image was obtained by adopting the minimum values at each pixel from all images. The final images were obtained by subtracting the minimal image from the original C3 images. While the spacecraft attitude data is absent in the C3 FITS headers (Bill Thompson, private communication), the comet was bright enough to be tracked by a computer algorithm from image to image. An example of a processed image from C3 is shown in Figure (\ref{fig:hi1-cor2-c3}) on the bottom panel. \\

In Fig. (\ref{fig:hi1-cor2-c3}), projected comet paths are overlaid on each image from three different instruments for their respective observation periods. The comet moves from  right to left in all three instrument field-of-view.\\

\section{Comet Tails}\label{sec:tails}
\begin{figure}
\centering
\includegraphics[width=\columnwidth]{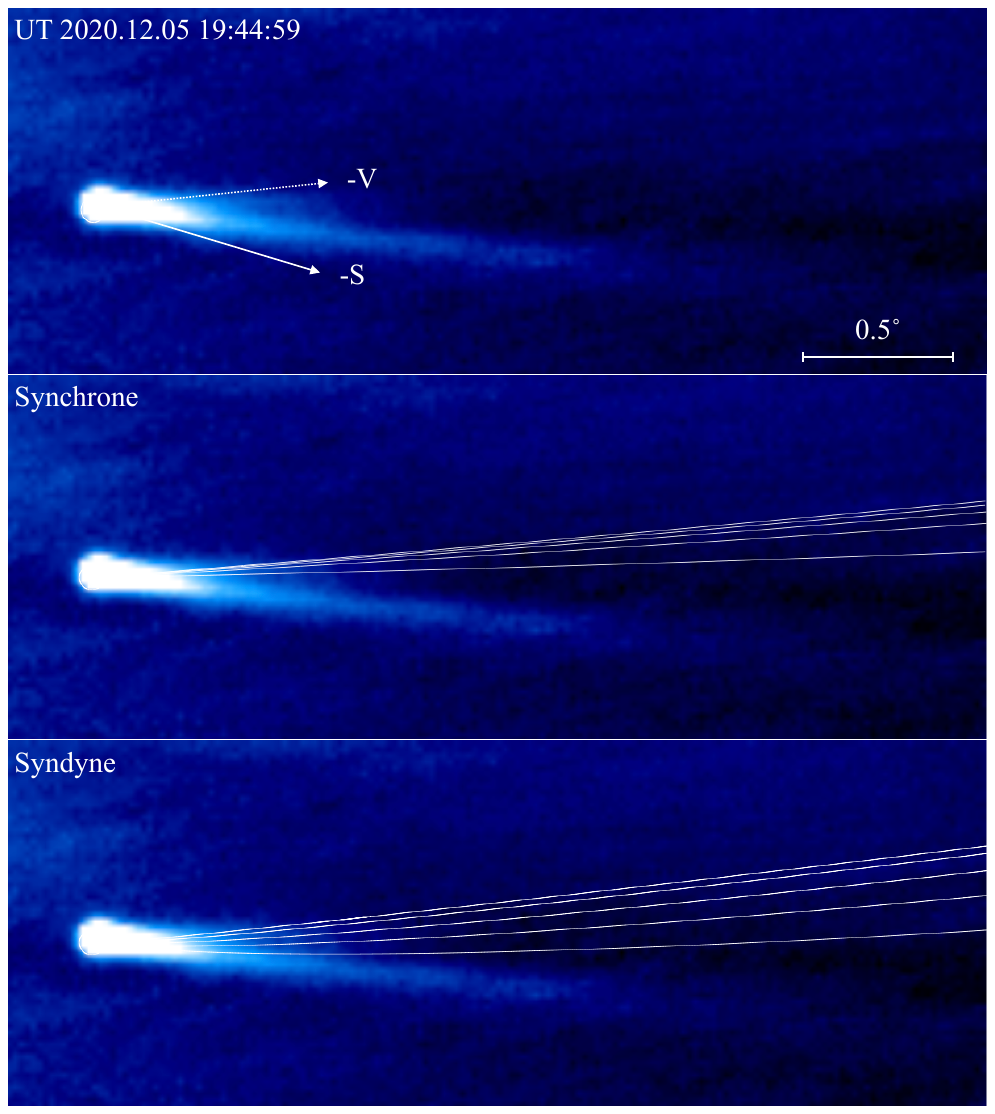}
\caption{An example image taken with the HI-1 wide field camera. Top: the median image from UT 2020 December 05.  The anti-Sun ($-S$) and anti-orbital  ($-V$) directions are indicated by solid and dashed arrows.  The bright long tail is the plasma tail. A faint tail above the plasma tail is the dust tail. The middle and bottom panels are the modeled synchrones and syndynes overlapping the image. Synchrones were computed for ejection times 20, 40, 60, 80, and 100 days prior to the date of observation (curves from bottom to top, in the figure). Syndynes were computed for $\beta$ = 1, 0.3, 0.1, 0.03, 0.01 (curves from bottom to top), corresponding to nominal particle radii 1, 3, 10, 30 and 100 $\mu$m, respectively. Celestial north is up,  east is to the left. The region shown is 3.2\arcdeg$\times$1.2\arcdeg~(160$\times$60 pixels). \label{fig:comet-hi1}}
\end{figure}

To enhance the signal to noise levels, we computed image medians using data taken over time spans between 0.2 and 1.0 days (Tables \ref{tbl:hi1}, \ref{tbl:cor2}, and \ref{tbl:c3}). All three tables list the dates and times of the image medians, position angles of anti-sun ($\theta_{-S}$) and anti-orbital-motion ($\theta_{-V}$) directions, of the measured position angles of dust ($\theta_{dust}$) and plasma ($\theta_{plasma}$) tails. Although dust is not our primary interest, we computed syndyne and synchrone dust models to compare with the data, assuming a range of particle sizes and ejection dates.  In these models, dust particles are released at zero velocity and accelerated by the gravity of the Sun and by radiation pressure. The particle  size is described by a dimensionless parameter, $\beta$, equal to the ratio of the acceleration induced by radiation pressure to that owing to local solar gravity \citep{Bohren83}.  To a first approximation, $\beta \sim a^{-1}$,  where $a$ is the particle radius in microns. These models not only reveal an estimate of the effective particle sizes in the dust tail, they also serve the purpose of identifying and separating dust from plasma tails in the plane of sky. The synchrone and syndyne curves are superimposed on the comet image in figures (\ref{fig:comet-hi1}), (\ref{fig:comet-cor2}) and (\ref{fig:comet-c3}). Synchrones were computed for ejection times 20, 40, 60, 80, and 100 days prior to the date of observation (curves from bottom to top, in the figure). Syndynes were computed for $\beta$ = 1, 0.3, 0.1, 0.03, 0.01 (curves from bottom to top), corresponding to nominal particle radii 1, 3, 10, 30 and 100 $\mu$m, respectively. \\

\subsection{Observations of Dust and Plasma Tails}
\begin{figure}
\includegraphics[width=\textwidth]{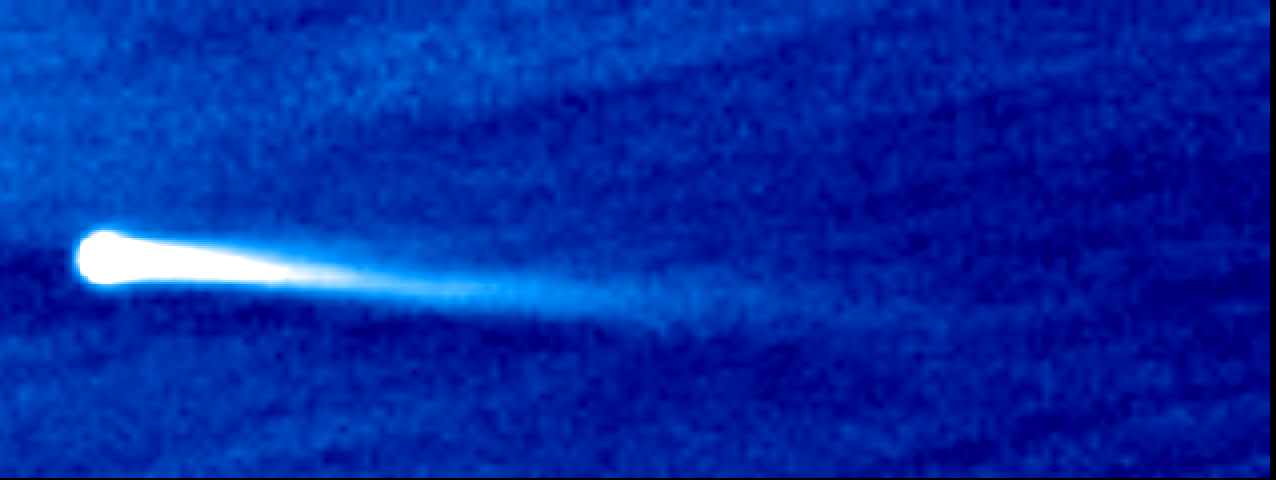}
\caption{S3 in the HI-1 camera. This is one frame from the animated series shown in the electronic edition. The bright tail is plasma.  The faint, approximately horizontal structure above the plasma tail is dust.  Image size is 3.2\arcdeg$\times$1.2\arcdeg (160$\times$60 pixels).\label{fig:animations-hi1}}
\end{figure}

{\bf HI-1:} A total of 26 HI-1  median images were generated as listed in Table (\ref{tbl:hi1}). An example median image, from UT 2020 December 5, is shown in the top panel of Figure (\ref{fig:comet-hi1}).  The STEREO-A images are aligned solar north up (Karl Battams, private communication). Before combination into image medians, we rotated the individual images  to equatorial orientation (North up, East left) using the telescope header roll angles.  An animation of the median images (Figure \ref{fig:animations-hi1}) shows a bright tail growing longer and brighter as the comet approaches perihelion. The tail position angles are broadly consistent with the antisolar ($-S$) direction. Above the bright long tail, a faint tail branch is seen having position angle closer to the negative heliocentric velocity (-V) direction. \\

Both synchrone and syndyne trajectories  have  position angles  $\theta_{PA} > 270\degr$ showing that  the faint and short upper component  in the HI-1 image is  the dust tail with characteristic dust  particle sizes  $\sim10 \mu$m. The lower and brighter long tail is the plasma tail.   \\

\begin{figure}
\centering
\includegraphics[width=\columnwidth]{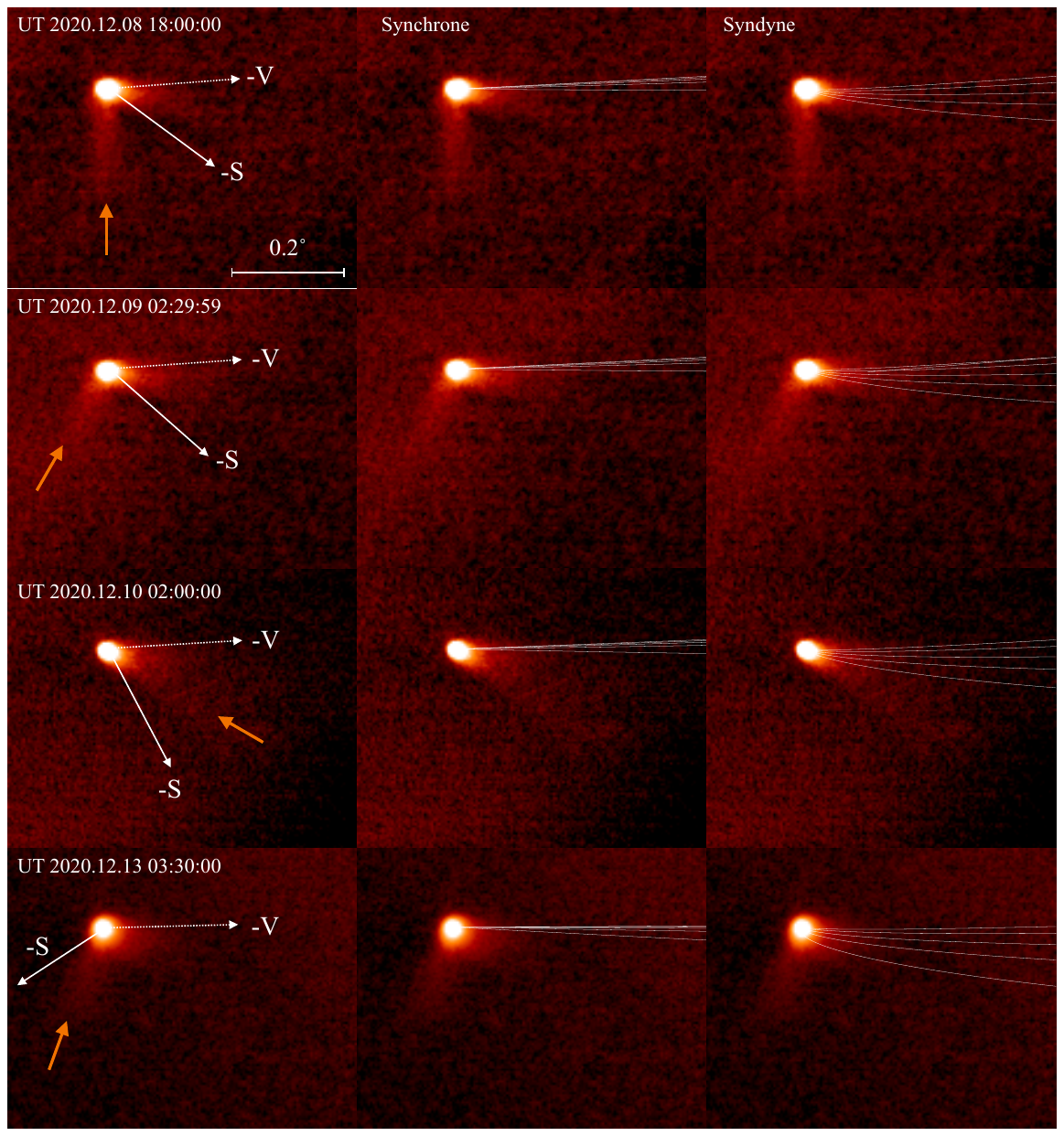}
\caption{Left Column: Examples of median COR2 images showing the rapidly changing position angle of the plasma tail (indicated by orange arrows). The  anti-Sun ($-S$) and anti-orbital velocity ($-V$) directions are indicated by solid and dotted arrows.  Middle and Right columns: The superimposed loci of synchrones and syndynes computed with same parameters as  in Fig.~(\ref{fig:comet-hi1}). Celestial north is up and  east  to the left. The region shown is  $\sim$0.6\arcdeg$\times$0.5\arcdeg~(150$\times$120 pixels).
\label{fig:comet-cor2}}
\end{figure}

\begin{figure}
\centering
\includegraphics[width=0.7\columnwidth]{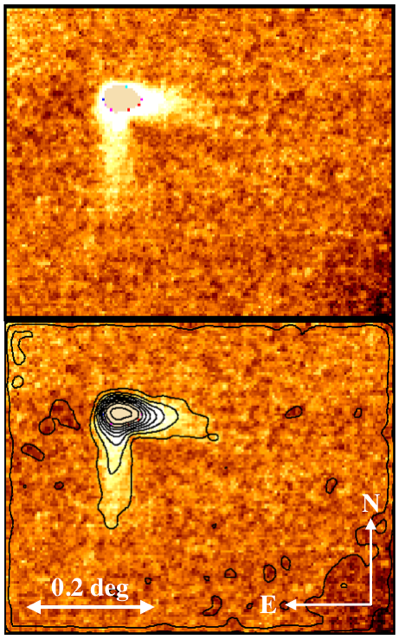}
\caption{COR2 camera image shown raw (upper) and contoured (lower). The plasma tail points to the south, the dust tail to the west, in this image. An animation including this image is available in the electronic edition. Image size is 0.6\arcdeg$\times$0.5\arcdeg (150$\times$120 pixels).\label{fig:animations-cor2}}
\end{figure}

{\bf COR2} Seventeen median-combined images were generated from COR2 data, as summarized in Table (\ref{tbl:cor2}). The left column of Figure (\ref{fig:comet-cor2}) displays four selected images, while the middle and right columns show overplotted synchrone and syndyne models. In all images, except for the one from UT 2020 December 10, two tails can be clearly seen, separated from each other. The syndyne dust models fit better to the thick tail profiles compared to the synchrones, suggesting that the dust particles are distributed over a range of sizes, which is consistent with the HI-1 data. Notably, the tail observed around position angle $270\degr$ is identified as the dust tail, while the second tail with variable position angles is attributed to the plasma tail. One remarkable finding from the COR2 image sequence is that, while the position angle of the dust tail remains stable near $\theta_{PA}=270\degr$, the plasma tail (indicated by orange arrows in Figure \ref{fig:comet-cor2}) exhibits unprecedented and rapid swings around the antisolar direction. This tail swing is prominently captured in the animation shown in Figure (\ref{fig:animations-cor2}) and discussed in Section 4. \\

On December 10, the two tails of the comet appear merged, making it difficult to distinguish one from the other. This merging of the tails lasted from UT 12/10 02:00 to 12/12 11:15. During this time, the plasma tail is completely invisible at UT 12/10 23:29 and 12/11 06:45. The position angles of the plasma tail at these times are omitted from Table (\ref{tbl:cor2}). There are several reasons why the plasma tail may be obscured from observers,  while we do recognize that the dust tail is located at $\theta_{PA} \approx 270^\circ$.  The small phase angles (see Table (\ref{tbl:cor2})) and the alignment of the tail with the sun-comet line may contribute to the plasma tail's obscurity. The strong background emission from the solar corona also plays a role in masking the plasma tail. \\

\begin{figure}
\centering
\includegraphics[width=0.9\columnwidth]{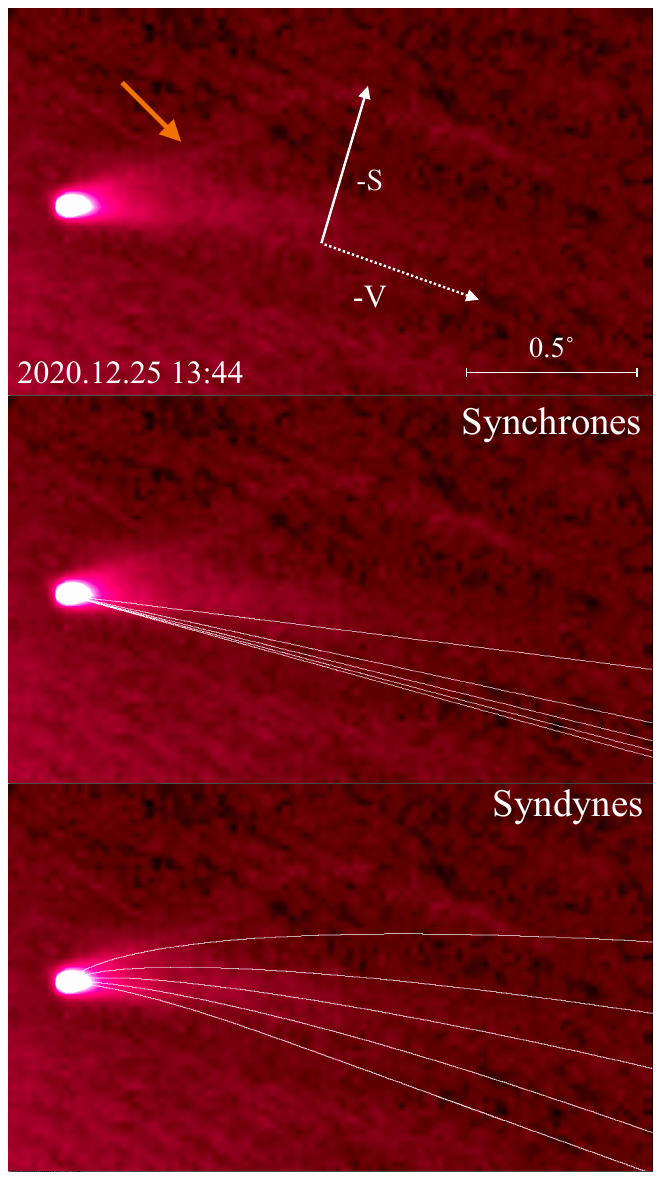}
\caption{Top: an example of the median image of LASCO/C3. -V (dotted arrow) and -S (solid arrow) represent anti-motion and anti-Sun directions at the location of the comet.  The plasma tail is indicated by an orange arrow. Middle and bottom: the computed synchrone and syndyne models superimposed on the comet image to simulate the dust tails. The solid white curves are the modeled synchrones and syndynes calculated with the same parameters as those in Figures (\ref{fig:comet-hi1}). North is up, east to the left and the region shown is  $\sim$1.9\arcdeg$\times$1.6\arcdeg~(120$\times$100 pixels). \label{fig:comet-c3}}
\end{figure}

{\bf LASCO/C3:} The telescope roll angles are not stored in the LASCO FITS headers and therefore, in order to align the C3 images with the celestial north, we measured the angles using the positions of known stars in the C3 images. In total, 25 background suppressed median images  (see Table (\ref{tbl:c3})) were generated from LASCO/C3 data. An example is shown in the top panel of Figure (\ref{fig:comet-c3}). The synchrone (second panel) and syndyne (third panel) models in Figure (\ref{fig:comet-c3}) reveal the wide large dust tail around the position angle $270\degr$. At smaller particle sizes of about $<30\mu$m, syndynes fit to the dust tail better than synchrones, but the synchrone with  longer ejection times (e.g.~100 days) seems to match better to the dust tail. Meanwhile, the plasma tail, indicated by an orange arrow, is well separated from the dust tail.  \\

\subsection{Tail Properties}
\begin{figure}
\centering
\includegraphics[width=\textwidth]{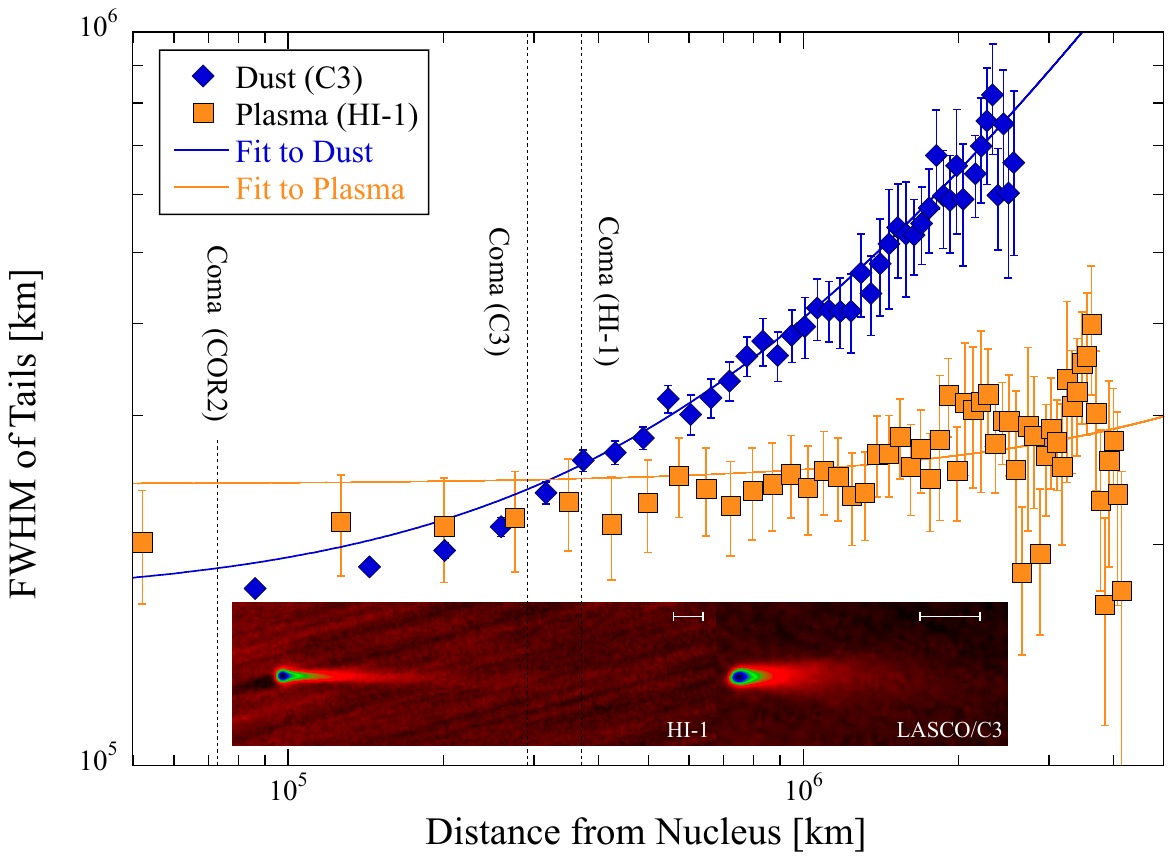}
\caption{Full width at Half Maximum (FWHM) of the comet tails as function of distance from nucleus. The error bars are the 3$\sigma$ uncertainties of the tail width computed from the Gaussian fits. The coma positions are marked with vertical dotted lines for each camera.  The embedded images are composites of HI-1 and LASCO/C3 images from which the widths are measured (dust widths are from LASCO/C3 and plasma widths are from HI-1).  Scale bars on the images are 1 Mkm at the comet.  \label{fig:tails-fwhm}}
\end{figure}

{\bf Tail Widths:}
Tail width measurements support the identification of the dust and plasma tails.  At each position along the tail we measured the Full Width at Half Maximum (FWHM, $w$) defined as the perpendicular width at half the maximum surface brightness. Figure (\ref{fig:tails-fwhm}) shows $w$ as a function of distance from the nucleus, with errors  estimated by comparing with Gaussian profiles. The nucleus FWHM is about 5 pixels for COR2 and 3 pixels for HI-1 and C3, reflecting the different point-spread functions of the instruments. The HI-1 and C3 cameras provide greater sensitivity to low surface brightness emission against the sky background. We obtain a composite dust tail image with LASCO/C3 images, and a composite plasma tail image with HI-1 images. They are embedded in Figure (\ref{fig:tails-fwhm}). As expected from the dynamics of dust, a linear fit  shows that the dust tail is strongly flared, with width  $w=(0.24\pm 0.02)r+(1.68\pm 0.24)\times10^5$ [km], where $r$ is the distance from the nucleus in km.  By comparison, the width of the plasma tail varies only slightly with increasing distance from nucleus, consistent with a magnetically confined structure. The fit to the plasma tail yielded $w=(0.01\pm0.01)r+(2.42\pm0.14)\times10^5$ [km].   \\

{\bf Tail Lengths:}
The tails fade progressively to invisibility with increasing distance from the nucleus allowing only lower limits to the lengths to be placed.  A rough estimate of the length was obtained using the DS9 display\footnote{http://ds9.si.edu/doc/ref/how.html} with the {\it histogram equalization} function on the best tail image. This technique increases global contrast of images which have limited range of values. The longest dust tail is found in C3 camera images at 6.5 Mkm. The plasma tail length is given by the HI-1 camera at 10.0 Mkm. The  COR2 images are less sensitive to low surface brightness emission and reveal the dust and plasma tails only to $\sim$0.7 Mkm.\\

{\bf Tail Position Angles:}
The position angles of the dust and plasma tails ($\theta_{dust}$, $\theta_{plasma}$) are crucial quantities to this study, especially $\theta_{plasma}$.  We tried different methods to measure the tail directions, including Gaussian fits to the tail profiles made perpendicular to their axes. In the end, the tail position angles were determined manually by fitting a straight line to data points along the axes with uncertainties determined from multiple measurements. The position angles of the tails were measured counterclockwise from the celestial north (see images shown as examples in Figures (\ref{fig:comet-hi1}, \ref{fig:comet-cor2}, and \ref{fig:comet-c3})). Figure (\ref{fig:pa-tails}) shows the  position angles as functions of time, with three panels to represent the observations from wide-field camera HI-1 (top),  narrow-field camera COR2 (middle)  and  LASCO/C3 (bottom). Blue and orange circles distinguish the position angles of the dust and plasma tails. The solid blue and red curves represent the position angles of -V and -S viewed from each telescope. The measured position angles and their uncertainties are also listed in Tables (\ref{tbl:hi1}, \ref{tbl:cor2} and \ref{tbl:c3}).   \\

\section{Discussion}\label{sec:discussion}

The discussion will be focused on the plasma tail. Position angles of the tail are used to infer the solar wind speed, and other conditions.\\

\subsection{Determination of the Plasma Tail Aberration Angle}
\begin{figure}
\centering
\includegraphics[width=\textwidth]{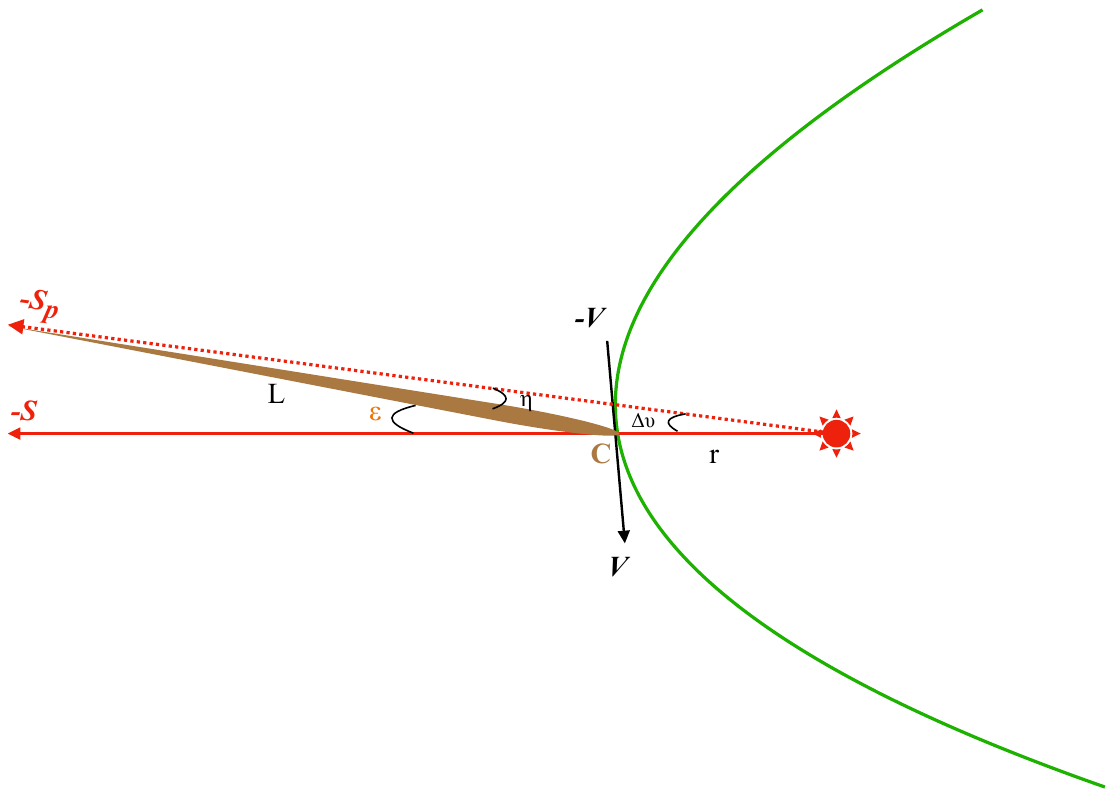}
\caption{Comet orbit (green arc) viewed from above. r: comet heliocentric distance; {\bf V}: comet orbital velocity; {\bf -S}: antisolar direction; {\bf -S}$_p$: the antisolar direction corresponding to the comet aberration; $\varepsilon$: aberration angle; $\Delta\nu=\nu-\nu_p$ ($\nu$: true anomaly at the observing time; $\nu_p$: true anomaly at the time associated with the observed tail position angle); {\bf L}: the plasma tail length; and $\eta$: the angle between {\bf -S}$_p$ and the plasma tail axis. The relation $\varepsilon=\eta+\Delta\nu$ is independent of tail length. Taking $\eta\rightarrow 0$, $\varepsilon\sim \Delta \nu$.\label{fig:orbit-view}}
\end{figure}

The comet aberration angle is measured in the comet orbital plane (Figure \ref{fig:orbit-view}). At the comet position, C, a number of quantities are plotted with respect to the Sun: the cometary velocity {\bf V} (and the anti-motion-velocity {\bf -V}); anti-Sun direction {\bf -S}. The comet and its plasma tail are represented by a brown tail-like image. It extends to the opposite direction of {\bf V} forming an aberration angle $\varepsilon$ with respect to the {\bf -S} direction (sun-comet line). The tail intercepts with another anti-sun direction {\bf -S$_p$} marking the sun-comet line  at an earlier time. As the comet moves along its orbit, its progression can be measured with the comet true anomaly which is the angle in the orbit plane from the perihelion direction to the comet, measured positively in the direction of motion. At the time of observation ($t$), the comet true anomaly is $\nu$. But the comet plasma tail tilts backward opposite of the motion direction, corresponding to the true anomaly $\nu_p$ at an earlier time ($t_p$). As a result, $t-t_p>0$, and $\Delta \nu=\nu-\nu_p>0$. Under assumption that the cometary tails  lie in the orbital plane, we have a triangle of the sun-comet-intercept point of the comet tail with sun-comet line in the {\bf-S$_p$} direction. In this triangle, {\bf r} is the heliocentric distance of the comet;  {\bf L} is the length of the comet tail.  The comet aberration angle is simply calculated as $\varepsilon=\Delta \nu+\eta$. We note that this relation is independent of the tail length, {\bf L}, regardless how $\eta$ and $\Delta\nu$ change in quantity.  Taking $\eta\rightarrow 0$, we can reasonably put $\varepsilon\sim \Delta\nu$. \\

\begin{figure}
\centering
\includegraphics[width=0.85\textwidth]{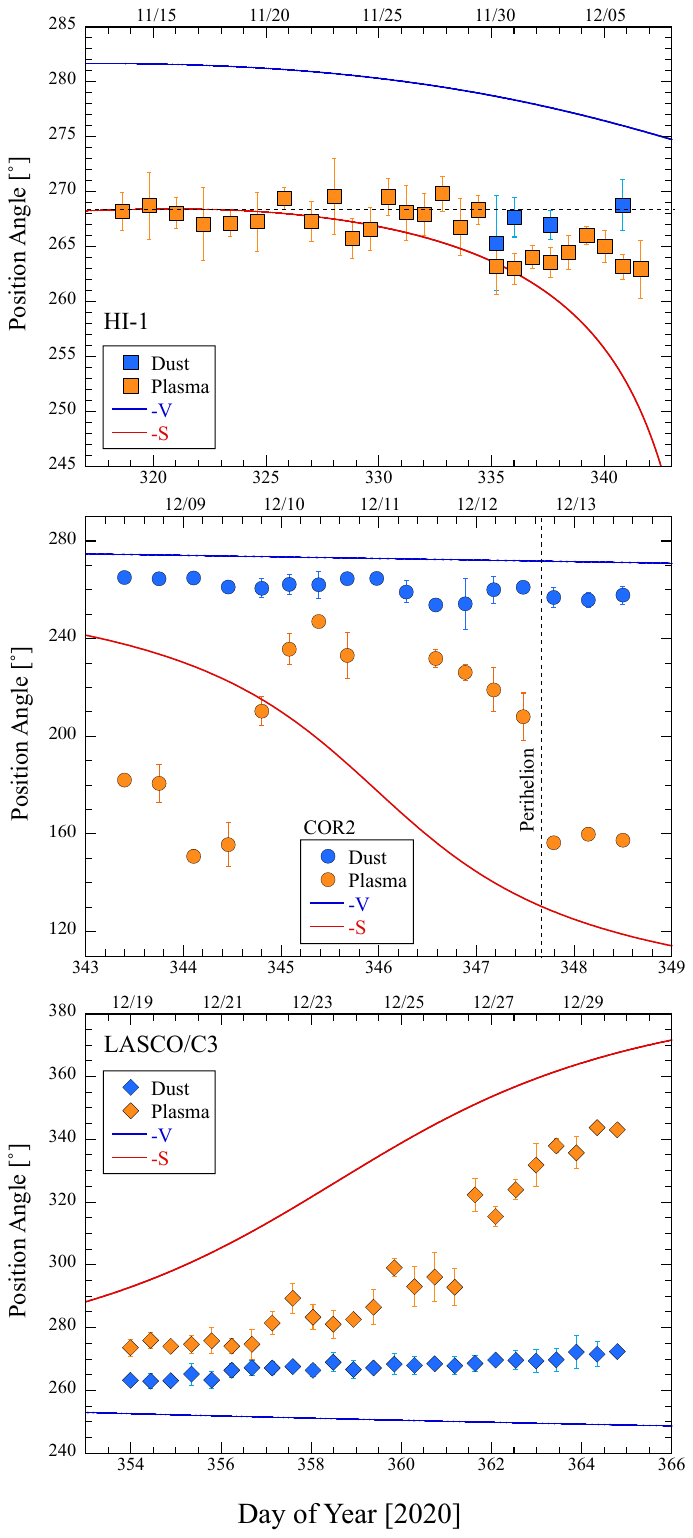}
\caption{Time dependence of the position angle of  the dust (blue) and plasma (orange) tails from HI-1 (top), COR2 (middle) and C3 (bottom). Position angle uncertainties on some points are comparable to the size of the symbols used to plot the data. Red and blue solid curves are the anti-Sun ($-S$) and anti-orbital-motion ($-V$) directions, respectively. 
The time of perihelion is marked with a vertical dotted line on the COR2 (middle) panel. The data points are presented in Tables (\ref{tbl:hi1}, \ref{tbl:cor2} and \ref{tbl:c3}). Note that five orange data points are above the horizontal dotted line on top panel. The comet aberration angles cannot be resolved on these data points using the current technique. The calendar dates are shown on the top horizontal axis. \label{fig:pa-tails}
}
\end{figure}

\begin{figure}
\centering
\includegraphics[width=0.8\textwidth]{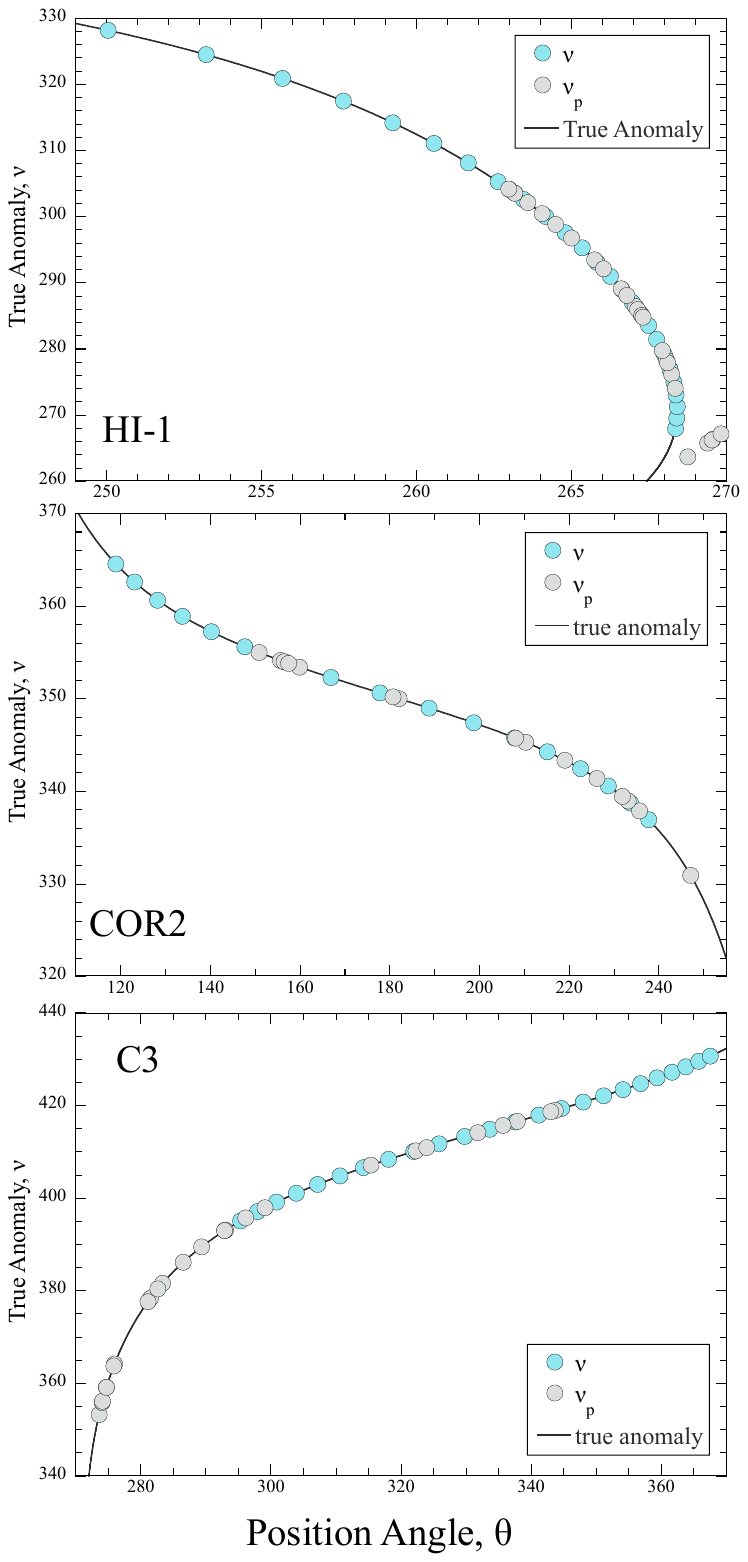}
\caption{The comet true anomaly ($\nu$) as function of anti-solar position angle ($\theta$). Functions are plotted as solid black curves provided by HORIZONS in all three panels. Light blue circles represent comet true anomalies ($\nu$) at the times of observations. Grey circles represent the comet true anomalies ($\nu_p$) inferred from the tail position angles ($\theta_{plasma}$). Three panels correspond to HI-1, COR2 and C3 instruments. Note the five grey data points are not on the black curve on top panel with HI-1. The true anomalies cannot be solved for these data points. \label{fig:tru_anom}}
\end{figure}

To determine $\Delta \nu$, we use the relation between the true anomaly ($\nu$) and the position angle of the anti-Sun direction  ({\bf -S}) projected to the observer's plane of sky ($\theta_{-S}$). The relation $\nu(t)=f(\theta_{-S}(t))$ is established by the quantities  ``Tru\_Anom'' and ``PsAng'' in the JPL HORIZONS' ``Observer Table''.  The observed position angle of the plasma tail ($\theta_{plasma}$) often differs from -S direction ($\theta_{-S}$) (see Figure (\ref{fig:pa-tails})), which results in a different true anomaly from a different time $\nu_p=f(\theta_{plasma})$ on the orbital plane. Figure (\ref{fig:tru_anom}) demonstrates the comet true anomalies $\nu$ (light blue circles) at the observing times and the true anomalies $\nu_p$ (grey circles) corresponding to the observed tail position angles at different times. Both $\nu$ and $\nu_p$ fall along the $\nu(t)=f(\theta_{-S}(t))$ function curve (black solid curve). On the top panel with HI-1, five grey data points are off the function curve. The aberration angles cannot be resolved for these data points. This underscores the limitation of the technique. As a result, these five data points are left with $\varepsilon$ listed in Table (\ref{tbl:hi1}). They also appear above the dotted horizontal line on the top panel (HI-1) of the Figure (\ref{fig:pa-tails}).\\

\begin{figure}
\centering
\includegraphics[width=0.9\textwidth]{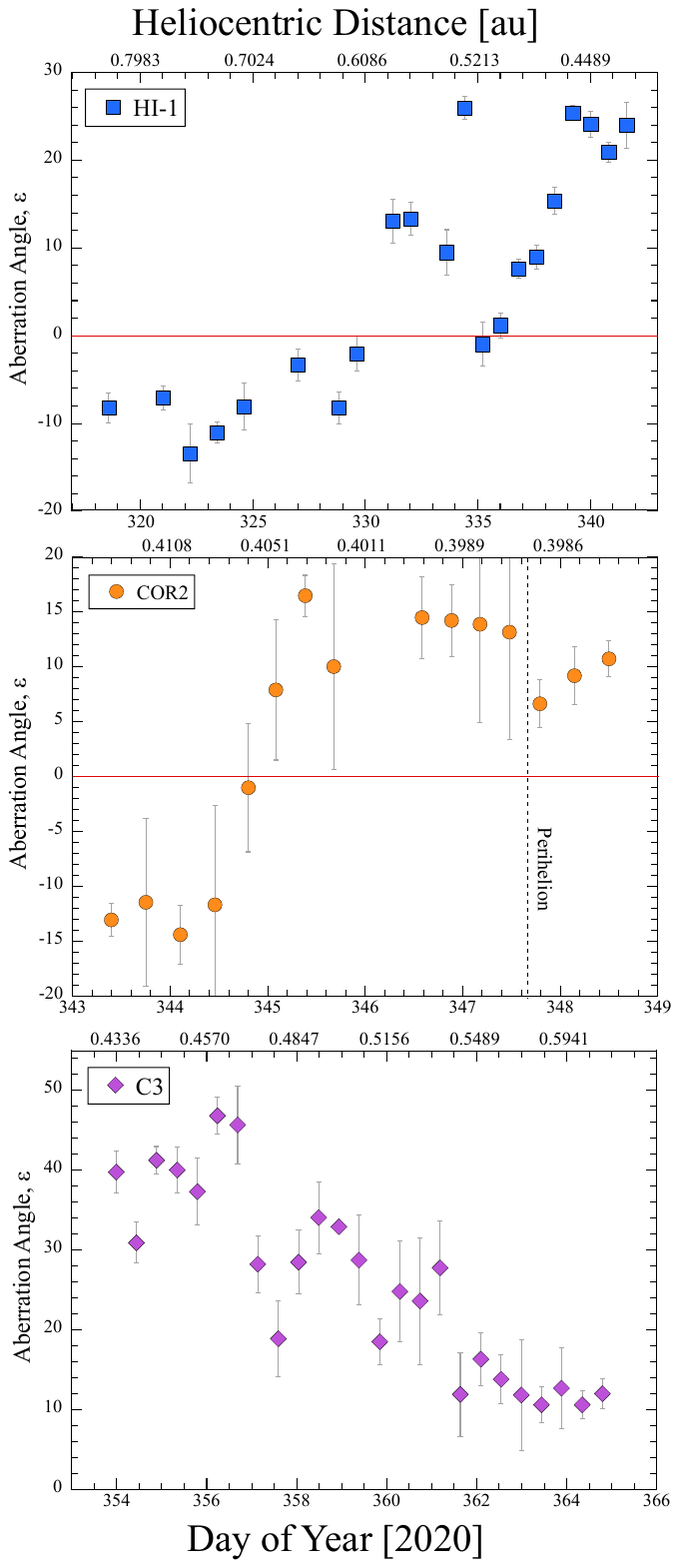}
\caption{The comet plasma tail aberration angles as function of time (bottom horizontal axis) and the heliocentric distance (top horizontal axis) on the comet orbital plane.  Three panels are the measurements from three instruments, HI-1 (top), COR2 (middle) and C3 (bottom).  Red horizontal lines show  the sun-comet line, $\varepsilon=0\degr$. \label{fig:aberration-obs}}
\end{figure}

Figure (\ref{fig:aberration-obs}) shows the aberration angles ($\varepsilon$) derived from observed position angles of the plasma tail ($\theta_{plasma}$) as function of time. Those errors in the observed position angles (see Figure (\ref{fig:pa-tails})) are adapted for the errors of the observed aberration angles, $\Delta\varepsilon$. The aberration angles also are found in the column $\varepsilon$ in Tables (\ref{tbl:hi1}, \ref{tbl:cor2} and \ref{tbl:c3}). For those data points with $\varepsilon\sim\Delta \nu=\nu-\nu_p >0$, the indication is that the tail tilts away from the comet motion; For those data points where $\varepsilon<0$, the tail is tilted forward in the same direction as the orbital motion ({\bf V}). \\

\subsection{Solar Wind Velocity Inferred from Plasma Tail Position Angles}
From the equation (\ref{eq:alfven}), the solar wind radial velocity can be inferred to 
\begin{equation}
w_r=\frac{v_t}{\tan\varepsilon}+v_r
\label{eq:swv}
\end{equation}

The errors are estimated from:
\begin{equation}
\Delta w_r=\left|-\frac{v_t}{\sin^2\varepsilon}\right|\Delta\varepsilon
\end{equation}
where $\Delta\varepsilon$ is in radians. Units for $w_r$, $v_t$ and $\Delta w_r$ are km/s.\\

\begin{figure*}
\includegraphics[width=0.9\columnwidth]{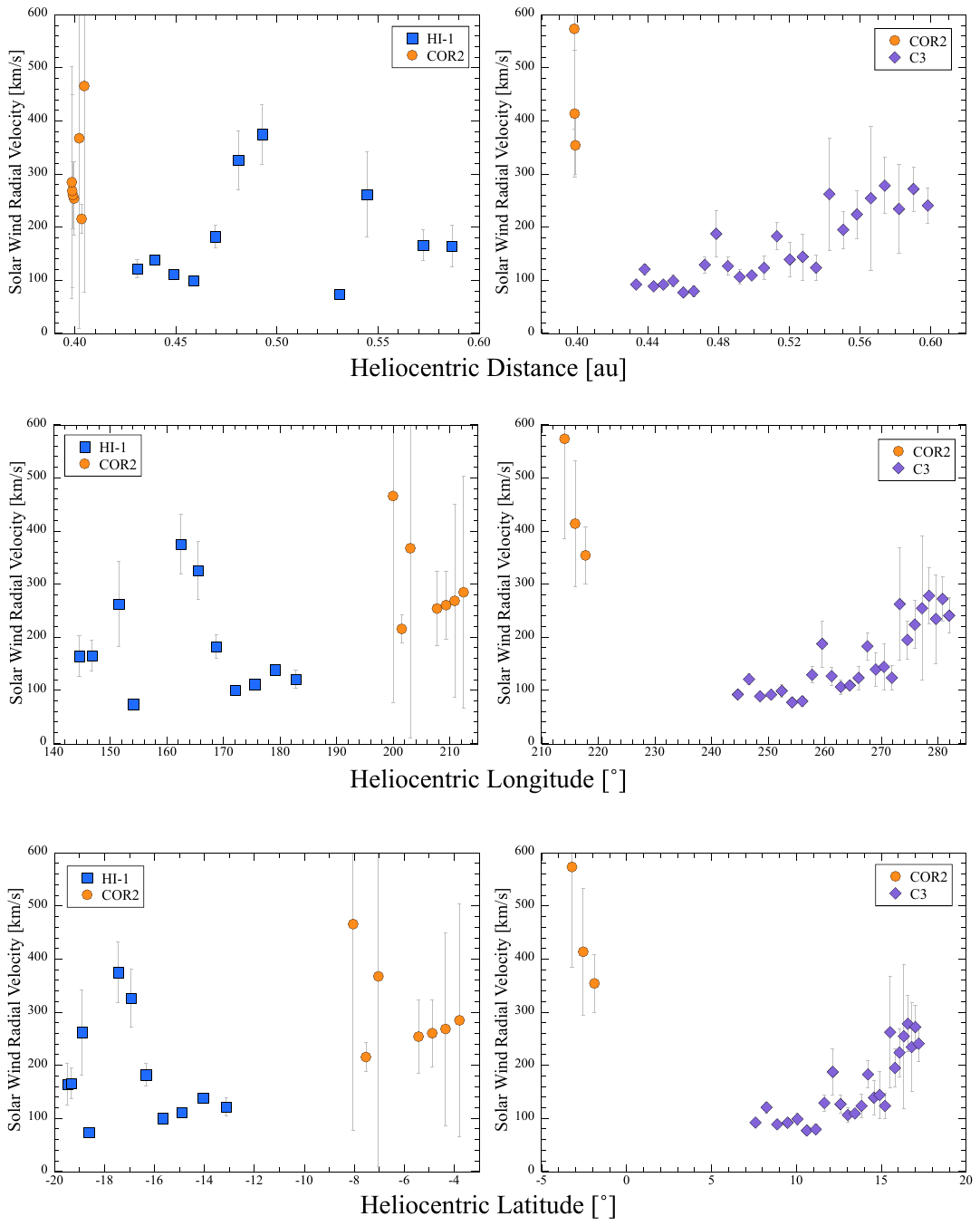}
\caption{Inferred solar wind radial velocities as functions of heliocentric distance (top), longitude (middle), and latitude (bottom). Three plots on the left represent the solar wind velocities as S3 moves inbound. Those on the right represent the solar wind as S3 moves outbound. Symbols distinguish measurements  made with  HI-1 (blue squares), COR2 (orange circles), and LASCO/C3 (purple diamonds). \label{fig:swv}}
\end{figure*}

The cometary radial velocity $v_r$ and the absolute velocity $v$ are known quantities provided by JPL HORIZONS. The tangential component is $v_t=\pm\sqrt{v^2-v_r^2}$, where $v_t>0$ as it points at the comet motion direction. The inferred solar wind velocities and their error bars are shown  in Figure (\ref{fig:swv}) as functions of heliocentric distance (top), longitude (middle) and latitude (bottom). They are found in ``$w_r$'' columns of Tables (\ref{tbl:hi1}, \ref{tbl:cor2} and \ref{tbl:c3}). The data points with $\varepsilon<0$ are not presented with the values of $w_r$ because such aberrations are probably caused by other phenomena rather than the radial solar wind, as we will discuss in the following sections. Some data points have very large uncertainties of the solar wind velocity. These data points are often coincident with the aberration angle $\varepsilon\la2\degr$. This is specially evident with observations of HI-1 (see Table \ref{tbl:hi1}).\\

Figure (\ref{fig:swv}) consists of two sets of plots, with the left side showing the solar wind radial velocities as a function of heliocentric distance, longitude, and latitude as comet C/2020 S3 (Erasmus) moves inbound towards the Sun. The right side, on the other hand, shows the same quantities but as the comet moves outbound away from the Sun. Comet S3 appears to be in the slow solar wind velocity domain during observations, with wind speeds approximately doubling as the comet travels inbound compared to its outbound journey. \\

To verify the solar wind environment around comet S3, we compare the solar wind measurements with large-scale, physics-based space weather prediction models known as WSA-Enlil Solar Wind Prediction models: Wang-Sheeley-Arge (WSA)-Enlil Solar Wind Prediction. These models are generated at The Space Weather Prediction Center (SWPC)\footnote{https://www.swpc.noaa.gov/products\/wsa-enlil-solar-wind-prediction}, which provides advance warning of solar wind structures and Earth-directed coronal mass ejections (CMEs). \\

We downloaded the models generated by SWPC for the time period from mid-November to the end of December 2020. We examine models of the solar wind velocity structures in the ecliptic plane, looking down from above the solar north. The positions of Earth, STEREO-A, and STEREO-B are also given in the simulations, allowing for a comparison between the model results and the actual observations made by STEREO-A and SoHO. The models predict solar wind velocities $<$550 km/s during the HI-1 observations;  $<$400 km/s during the COR2 observations; and $<$300 km/s during the C3 observations. The models have the minimum velocity 200 km/s, while the observed solar wind velocities are often $<$200 km/s around the S3 on the comet's journey both inbound and outbounds (Fig. \ref{fig:swv}). One possibility is that the solar wind mass loading due to pick up of cometary ions may decelerate the solar wind \citep{1987JGR....9213409O}. The evidence is that the solar wind velocities are $<$ 100 km/s as S3  near the perihelion ($<$0.455 au) with the exception of  observations taken in the COR2 visibility window. \\

\subsection{Plasma Tail Influenced by Non-Radial Solar Wind Flow\label{sec:nonr}}
On the top panel of Figure (\ref{fig:aberration-obs}), some data points have $\varepsilon<0$, meaning that the tails tilt forward in the cometary motion direction opposite to the non-radial flows carried by the solar rotation \citep{1967ApJ...148..217W}. This implies the presence of the azimuthal  solar wind flow against  Archimedes spiral which is about a few km/s according to an analytical models by \citet{2016JGRA..121.4966T}. The equation (\ref{eq:alfven}) can be slightly modified by replacing $v_t$ with $v_t+w_{nonr}$. Then the non-radial velocity component can be estimated with $\varepsilon$:
\begin{equation}
w_{nonr}=\tan\varepsilon(w_r-v_r)-v_t
\label{eq:nonr}
\end{equation}
where $v_t$ and $v_r$ are the cometary tangential and radial components of the orbital velocity; $w_r$ is the solar wind radial velocity; $\varepsilon$ is simply the angle of the plasma tail with respect to the solar radial direction in negative values. The azimuthal component of the solar wind due to Archimedes spiral is omitted in this estimation. Therefore, the equation (\ref{eq:nonr}) gives the lower limit of the non-radial solar wind component in the cometary motion direction. Given an average $w_r=183.6$ km s$^{-1}$ during the HI-1 observations,  we obtain the average non-radial velocity $\sim-$66.8 km s$^{-1}$ in the same period of time. The negative value represents the non-radial velocity direction against the co-rotating direction. The individual inferred $w_{nonr}$ corresponding to each data point are listed in the last column of Table (\ref{tbl:hi1}).\\

Several causes can result in  non-radial solar wind flows. A survey of non-radial flows during nearly five years of observation reveals that the average  non-radial flow speed is $\sim30$ km s$^{-1}$. Furthermore, about half of the non-radial flow events are associated with interplanetary coronal mass ejections (ICMEs) with average speeds exceeding 100 km/s \citep{2004AnGeo..22.4397O}. The cause of non-radial flow by a CME is the diversion of flow in the sheath region behind an ICME shock front due to the compression of the shockwave \citep{1987JGR....9212399G}.  Another possible cause of non-radial flow is in the co-rotating interaction regions (CIRs) which can play a role in deflecting the fast and slow solar wind flows azimuthally in opposite directions  \citep{1999SSRv...89...21G}.  In the absence of the detectable ICMEs, we investigate whether CIRs are responsible for the observed non-radial flows during the time period under study (DOY318.6 to 329.6, as shown in the top panel of Figure \ref{fig:swv}). Enlil models are examined for clues regarding the presence of corotating streams. The models show that the corotating streams swept through the path of the S3 with velocities ranging from 200 to 550 km/s. However, without careful examination of the comet's path overlapping with the corotating regions, it is challenging to determine with certainty whether the non-radial flows are caused by CIRs, although this remains a possibility.\\

\subsection{Plasma Tail and ICME \label{sec:icme}}
The most noteworthy feature in our data is a rapid change in the plasma tail position angles occurring as soon as the comet enters the COR2 FOV. The plasma tail position angles  swing like a pendulum by an amount that is very large compared to the uncertainties of measurement (see the first 5 data points in the middle panel of Figure \ref{fig:pa-tails}).  The tail position angle began at  $\theta_{plasma}\sim 180\degr$ in the first two data points, then decreased to $\sim150\degr$ and then abruptly swung near the anti-solar direction ($\sim210\degr$). Despite the dramatic change in position angles in the observer's frame, the tail angles with respect to the Sun-comet line show a more modest variation. The average $-\varepsilon\sim -12.7\degr\pm3.5\degr$ in the comet orbital plane (see middle panel of Figure (\ref{fig:aberration-obs})) among the first four data points (a time period $\sim25.5$ hours). The tail angle then swings back over the next $\sim$8.5 hours to near the sun-comet line ($0\degr$) as it is shown with the 5th data point in the figure, and thereafter to $\sim$+15\degr. The abnormal tail orientation lasted at least $\sim34$ hours. \\

We searched for evidence to support the conjecture that the plasma tail of S3 might have been deflected by a coronal mass ejection.  While there are no in-situ observations to confirm an ICME interaction with comet S3, we do indeed find a Halo CME which occurred on UT 2020 December 7.68 from the NASA SoHO LASCO CME catalog\footnote{https://cdaw.gsfc.nasa.gov/CME\_list}. Using the Helioview\footnote{https://www.jhelioviewer.org} tool, we identify the likely  source of the CME as active region NOAA 12786.  This active region had rotated to the far side of the Sun from  STEREO-A at the time of the comet observation  (c.f. Figure \ref{fig:e-hi1-cor2}), meaning that the Halo CME travelled directly toward  the comet, which was also on the far side of the Sun from STEREO-A. Given the 0.418 au heliocentric distance  of S3 and an estimated  speed of 1407 [km s$^{-1}$]\footnote{https://cdaw.gsfc.nasa.gov/CME\_list/UNIVERSAL\_ver1/2020\_12/univ2020\_12.html},  the travel time  for the CME front to reach the comet is $\sim$12.4 hours, corresponding to arrival on UT December 8.2 (DOY 343.2). This arrival time approximately matches the  tail wagging motion detected in COR2 data (Figure \ref{fig:pa-tails}). Furthermore, the 34 hour time scale of the tail disturbance is comparable to the $\sim$30 hour average ICME crossing time \citep{2011SoPh..269..141W}. In the last column of Table (\ref{tbl:cor2}), we listed the non-radial wind velocity estimated with $\varepsilon<0$ using the equation (\ref{eq:nonr}). The average non-radial velocity is $>100$ km s$^{-1}$ quite consistent with the non-radial velocities caused by CMEs found in the survey by \citet{2004AnGeo..22.4397O}. While it is possible that the CME and the tail disturbance are only coincidentally related, we note that comparable CMEs are comparatively rare; only two such events occurred in the entire month of December.  In this interpretation, the absence of tail wagging in the earlier HI-1 and later C3 observations occurs simply because  no other ICME impacts  S3 during those  times. After passing an ICME, the tail angle should naturally return close to the solar wind radial  direction, as observed in Figure (\ref{fig:pa-tails}) in the middle panel.  We conclude that the plasma tail swing in COR2 is plausibly attributed to interaction with an ICME. \\


\section{SUMMARY}\label{sec:summary}

We present near-perihelion observations of long-period comet C/2020 S3 (Erasmus)  from the STEREO-A and SoHO spacecraft.  The STEREO-A camera HI-1 and the  coronagraph COR2 captured the comet from $\sim$ 0.836 au to perihelion at 0.398 au. About 4.3 days later, the comet was imaged by the LASCO/C3 coronagraph over the outbound heliocentric distance range $\sim$0.427 to 0.658 au. The comet moved from  heliocentric latitude $-20\degr$ to $20\degr$ and longitude 119$\degr$ to 285$\degr$ during our observations. In order to understand the comet plasma tail motions in the observer's plane of sky, a simple technique is developed to obtain the comet aberration angles ($\varepsilon$) from the observed position angles of the plasma tail. We summarize the observations and conclusions below:

\begin{itemize}
 \item We detected a large and transient excursion in the position angle of the plasma tail on UT 2020 December 8, from $180^\circ$ to $150\degr$ then back to $210\degr$, lasting about 34 hours.  The excursion cannot be caused by changes in the speed of the solar wind.  However, the excursion is coincident with the passage of an interplanetary coronal mass ejection (ICME).  We find that NOAA 12786 launched a halo CME with a direction and time consistent with the observed deflection of the comet tail.

\item We detected a long duration ($\gtrsim$11 days) wagging of the plasma tail about the Sun-comet line caused by non-radial flows. The Enlil solar wind models show that the corotating streams swept through the path of the comet with velocities ranging from 200 to 550 km/s. However,  careful examination of the comet's path overlapping with the corotating regions is needed to determine with certainty whether the non-radial flows are caused by CIRs, although this remains a possibility.

\item Wind speeds inferred from the  position angles range from $\sim$73.9 to $\sim$573.5 km s$^{-1}$, showing that comet S3 was in the slow solar wind region.  This is confirmed by Enlil models during the observation period. The average (median) solar wind radial velocities are 205.5 (182.3) km s$^{-1}$ as the comet moves in space. The  velocities show a very rough increase with decreasing heliocentric latitude in the STEREO-A observations. On the other hand, they tend to increase slightly with increasing heliocentric latitude in LASCO/C3. 

\item The average (median) plasma tail aberration angles are  $20.3\degr$ ($16.3\degr$) with minimum  and  maximum values $\varepsilon=1.2\degr$ and $\varepsilon=46.8\degr$, respectively. 

\item The technique to obtain the comet aberrations relies on  accurate measurements of position angles of the plasma tail in the equatorial frame. Therefore,  knowledge of the telescope roll angles between spacecraft image frames and the celestial coordinates is crucial. The technique is valid under the assumption that the comet tails lie in the orbital plane. The inferred solar wind radial velocity becomes unreliable when $\varepsilon<2^\circ$.

\item The dust tail is spatially distinguished from the plasma tail in the COR2 and LASCO/C3 cameras, but only  marginally resolved in HI-1 data. Comparison with synchrone and syndyne dust models indicates that the dust tail is dominated by particles with $\beta\sim0.1$ corresponding to characteristic sizes of order 10 $\mu$m. 

 \end{itemize}





\clearpage
\acknowledgments

We thank Drs. Yingdong Jia, John Raymond, Pedro Lacerda, Masateru Ishiguro and Bin Yang for reading the manuscript and providing comments. We would like to express our gratitude to the anonymous referee whose critical comments urged us to re-examine the observational data. We are grateful to Dr.~William Thompson for prompt help with  world coordinate system questions, and general data reduction. Dr. Bernhard Fleck  was also very kind to help with our inquiries. Dr. Karl Battams advised us regarding LASCO/C3 image rotations.

\clearpage
\begin{deluxetable}{llcc}
\tablecaption{Instrument Parameters\label{tbl:cameras}}
\tablehead{
\colhead{Instrument} & 
\colhead{Field of View}&
\colhead{Pixel Size}  &
\colhead{Bandpass [nm]} 
}
\startdata
HI-1/STEREO-A & 4\arcdeg-24\arcdeg (15-90 \RSUN)\tablenotemark{a}& 71.9\arcsec & 630-730 \\
COR2/STEREO-A &0.5\arcdeg-3.8\arcdeg~(2-15 \RSUN)\tablenotemark{b}& 14.7\arcsec & 650-750\\
C3/SoHO & 1\arcdeg-8\arcdeg(3.7-30 \RSUN)\tablenotemark{c}& 56.0\arcsec & 400-835\tablenotemark{d}
\enddata
\tablenotetext{a}{The wide field camera is aimed at an angle 14\arcdeg~off the solar limb from the principal axis of the instrument.}
\tablenotemark{b,c}{Both instruments are coronagraphs. The angles quoted are the inner and outer edges of the field of view, measured from Sun center.}
\tablenotetext{d}{The filter wheel contains  blue (420-520 nm), orange (540-640 nm), deep red (730-835 nm), infrared (860-1050 nm) and broadband clear (400-850 nm) filters. A narrow band H$_\alpha$ (2 nm) filter is mounted in a wheel holding three polarizers. However, the clear filter is normally used for the observations, which is the case for the current study.}
\end{deluxetable}

\clearpage

\begin{longrotatetable}
\begin{deluxetable}{lcccllrll}
\tablecaption{Measurements From HI-1\label{tbl:hi1}}
\tabletypesize{\footnotesize}
\tablehead{
\colhead{Date} & 
\colhead{Days\tablenotemark{a}} &
\colhead{$\theta_{-S}$\tablenotemark{b}}  &
\colhead{$\theta_{-V}$\tablenotemark{c}} &
\colhead{$\theta_{dust}$ \tablenotemark{d}} &
\colhead{$\theta_{plasma}$\tablenotemark{e}} &
\colhead{$\varepsilon$\tablenotemark{f}} &
\colhead{$w_r$\tablenotemark{g}} &
\colhead{$w_{nonr}$\tablenotemark{h} }
}
\startdata
2020-11-13 14:29& 1.0&268.3&281.6&\nodata&268.2$\pm1.7$ &-8.2&\nodata & -63.5\\
2020-11-14 19:15& 1.0&268.4&281.6&\nodata&268.7$\pm3.0$ & \nodata& \nodata &\nodata\\
2020-11-16 00:14& 1.0&268.4&281.5&\nodata&268.0$\pm1.3$ &-7.1&\nodata & -61.0\\
2020-11-17 05:00& 1.0&268.3&281.4&\nodata&267.0$\pm3.3$ &-13.4&\nodata & -86.8\\
2020-11-18 09:45& 1.0&268.3&281.3&\nodata&267.1$\pm1.2$ &-11.0&\nodata & -78.4\\
2020-11-19 14:29& 1.0&268.1&281.2&\nodata&267.2$\pm2.6$ &-8.1&\nodata & -68.3\\
2020-11-20 19:15& 1.0&267.9&281.1&\nodata&269.3$\pm1.0$ & \nodata& \nodata &\nodata\\
2020-11-22 00:14& 1.0&267.7&280.9&\nodata&267.3$\pm1.8$ &-3.3&\nodata & -52.6\\
2020-11-23 00:30&0.7&267.4&280.7&\nodata&269.5$\pm3.4$ & \nodata& \nodata&\nodata\\
2020-11-23 19:44&0.7&267.2&280.5&\nodata&265.7$\pm1.7$ &-8.2&\nodata & -73.3\\
2020-11-24 14:45&0.7&266.9&280.3&\nodata&266.6$\pm1.9$ &-2.0&\nodata & -50.8\\
2020-11-25 09:59&0.7&266.6&280.1&\nodata&269.5$\pm1.6$ & \nodata& \nodata &\nodata\\
2020-11-26 05:15&0.7&266.2&279.9&\nodata&268.1$\pm2.5$ & 13.0& 164.2$\pm 38.7$ &\nodata\\
2020-11-27 00:30&0.7&265.8&279.7&\nodata&267.9$\pm1.8$ & 13.3& 165.6$\pm 28.8$ &\nodata\\
2020-11-27 19:44&0.7&265.3&279.5&\nodata&269.8$\pm1.6$ &\nodata& \nodata &\nodata\\
2020-11-28 14:45&0.7&264.8&279.2&\nodata&266.7$\pm2.5$ & 9.5& 262.0$\pm 80.4$ &\nodata\\
2020-11-29 09:59&0.7&264.1&278.9&\nodata&268.3$\pm1.3$ & 25.9& 73.9$\pm 6.0$ &\nodata\\
2020-11-30 05:15&0.7&263.4&278.6&265.3$\pm$4.3&263.1$\pm2.4$ &-1.0&\nodata & -54.9\\
2020-12-01 00:30&0.7&262.6&278.3&267.6$\pm$1.7&263.0$\pm1.4$ & 1.2&\nodata&\nodata\\
2020-12-01 19:44&0.7&261.6&277.9&\nodata&264.0$\pm1.0$ & 7.7& 374.8$\pm 56.7$ &\nodata\\
2020-12-02 14:45&0.7&260.5&277.6&266.9$\pm$1.3&263.5$\pm1.3$ & 8.9& 325.9$\pm 54.7$ &\nodata\\
2020-12-03 09:59&0.7&259.2&277.2&\nodata&264.4$\pm1.5$ & 15.3& 182.1$\pm 21.9$ &\nodata\\
2020-12-04 05:15&0.7&257.6&276.8&\nodata&266.0$\pm1.0$ & 25.3& 99.8$\pm 4.7$ &\nodata\\
2020-12-05 00:30&0.7&255.6&276.4&\nodata&265.0$\pm1.4$ & 24.1& 111.2$\pm 9.1$ &\nodata\\
2020-12-05 19:44&0.7&253.1&275.9&268.7$\pm$2.3&263.1$\pm1.1$ & 20.9& 138.7$\pm 9.3$ &\nodata\\
2020-12-06 14:45&0.7&249.9&275.5&\nodata&262.9$\pm2.6$ & 23.9& 121.1$\pm 17.2$ &\nodata\\
\enddata
\tablenotetext{a}{Number of days used to make median images}
\tablenotetext{b}{Position angle of the  anti-Sun direction, degree, by HORIZONS}
\tablenotetext{c}{Position angle of the  negative heliocentric velocity, degree, by HORIZONS}
\tablenotetext{d}{Position angle of the dust tail and its uncertainty, degree.}
\tablenotetext{e}{Position angle of the plasma tail and its uncertainty, degree.}
\tablenotetext{f}{Aberration angle.}
\tablenotetext{g}{Inferred solar wind radial velocity  from  equation (\ref{eq:swv}). Unit:  km s$^{-1}$}
\tablenotetext{h}{Non-radial solar wind velocity from Equation (\ref{eq:nonr}) assuming $w_r=183.6$ in average. Unit: km s$^{-1}$. The nagative values represent the $w_{nonr}$ has in the opposite of the aberration.}
\end{deluxetable}
\end{longrotatetable}

\clearpage

\begin{deluxetable}{lclccllrll}
\tablecaption{Measurements From COR2\label{tbl:cor2}}
\tablehead{
\colhead{Date} & 
\colhead{Days\tablenotemark{a}} & 
\colhead{$\alpha$\tablenotemark{b}} &
\colhead{$\theta_{-S}$\tablenotemark{c}}  & 
\colhead{$\theta_{-V}$\tablenotemark{d}} & 
\colhead{$\theta_{Dust}$\tablenotemark{e}} & 
\colhead{$\theta_{Plasma}$\tablenotemark{f} } &
\colhead{$\varepsilon$\tablenotemark{g}} &
\colhead{$w_r$\tablenotemark{h}} &
\colhead{$w_{nonr}$\tablenotemark{i}}
}
\startdata
2020-12-08 09:30&0.3& 9.14&237.6&274.4&265.0$\pm$1.0&182.0$\pm1.5$ &-13.0&\nodata&-109.7\\
2020-12-08 18:00&0.3& 8.31&233.6&274.2&264.5$\pm$.71&180.7$\pm7.6$ &-11.4&\nodata& -104.1\\
2020-12-09 02:29&0.3& 7.52&228.6&274.0&264.8$\pm$2.0&150.8$\pm2.6$ &-14.4&\nodata&-114.8\\
2020-12-09 11:00&0.3& 6.77&222.4&273.8&261.1$\pm$1.6&155.6$\pm9.0$ &-11.6&\nodata&-105.2\\
2020-12-09 19:15&0.3& 6.12&215.0&273.6&260.6$\pm$3.9&210.3$\pm5.8$ &-1.0  &\nodata& -68.9\\
2020-12-10 02:00&0.2& 5.68&207.7&273.4&262.2$\pm$3.9&235.6$\pm6.3$ & 7.9  & 465.7$\pm 388.5$& \nodata\\
2020-12-10 09:14&0.2& 5.30&198.6&273.2&262.0$\pm$5.5&247.0$\pm1.8$ & 16.4& 215.4$\pm 26.7$& \nodata\\
2020-12-10 16:15&0.2& 5.08&188.7&273.0&264.5$\pm$2.0&233.1$\pm9.4$ & 10.0& 367.4$\pm 358.0$& \nodata\\
2020-12-10 23:29&0.2& 5.01&177.7&272.8&264.7$\pm$2.2& \nodata            & \nodata     & \nodata      &\nodata\\
2020-12-11 06:45&0.2& 5.13&166.8&272.6&259.1$\pm$4.5&  \nodata           & \nodata  & \nodata        &\nodata\\
2020-12-11 14:00&0.2& 5.41&156.7&272.4&253.8$\pm$1.8&231.9$\pm3.8$ & 14.5& 253.8$\pm 69.6$& \nodata\\
2020-12-11 21:14&0.2& 5.84&147.6&272.2&254.2$\pm$10.&226.1$\pm3.3$ & 14.2& 260.1$\pm 63.6$& \nodata\\
2020-12-12 04:15&0.2& 6.37&140.2&272.0&260.0$\pm$5.4&219.0$\pm8.9$ & 13.8& 268.2$\pm 181.6$& \nodata\\
2020-12-12 11:29&0.2& 7.01&133.8&271.8&261.0$\pm$2.2&208.0$\pm9.7$ & 13.2& 284.5$\pm 218.9$& \nodata\\
2020-12-12 18:59&0.3& 7.75&128.2&271.6&256.8$\pm$4.1&156.4$\pm2.1$ & 6.6& 573.4$\pm 188.7$& \nodata\\
2020-12-13 03:30&0.3& 8.64&123.1&271.4&255.8$\pm$3.1&159.8$\pm2.6$ & 9.2& 413.7$\pm 119.0$& \nodata\\
2020-12-13 12:00&0.3& 9.59&118.9&271.2&257.8$\pm$3.7&157.4$\pm1.6$ & 10.7& 353.9$\pm 54.3$& \nodata\\
\enddata
\tablenotetext{a}{Number of days used to make median images}
\tablenotetext{b}{S3 Phase Angle}
\tablenotetext{c}{Position angle of the  anti-Sun direction, degree.}
\tablenotetext{d}{Position angle of the negative velocity, degree.}
\tablenotetext{e}{Position angle of the dust tail, degree.}
\tablenotetext{f}{Position angle of the plasma tail, degree.}
\tablenotetext{g}{Aberration angle.}
\tablenotetext{h}{Inferred solar wind radial velocity with measured plasma tail position angles from the equation (\ref{eq:swv}). Unit:  km s$^{-1}$}
\tablenotetext{i}{Non-radial solar wind velocity, km s$^{-1}$. The nagative values represent the $w_{nonr}$ has in the opposite of the aberration.}
\end{deluxetable}

\clearpage

\begin{deluxetable}{lclcccrl}
\tablecaption{Measurements From LASCO/C3\label{tbl:c3}}
\tablehead{
\colhead{Date} & 
\colhead{Days\tablenotemark{a}} & 
\colhead{$\theta_{-S}$\tablenotemark{b}}  & 
\colhead{$\theta_{-V}$\tablenotemark{c}} & 
\colhead{$\theta_{Dust}$\tablenotemark{d}} & 
\colhead{$\theta_{Plasma}$\tablenotemark{e}} &
\colhead{$\varepsilon$\tablenotemark{f}} &
\colhead{$w_r$\tablenotemark{g}}
}
\startdata
2020-12-18 23:45&0.4&292.9&252.5&263.1$\pm$1.3&273.5$\pm2.6$ & 39.7& 91.9$\pm 6.9$\\
2020-12-19 10:30&0.4&295.3&252.4&262.9$\pm$2.3&275.9$\pm2.5$ & 30.8& 120.4$\pm 10.3$\\
2020-12-19 21:14&0.4&297.9&252.2&263.0$\pm$1.4&274.0$\pm1.7$ & 41.2& 88.5$\pm 4.1$\\
2020-12-20 08:15&0.4&300.8&252.0&265.2$\pm$3.4&274.6$\pm2.8$ & 40.0& 91.6$\pm 7.2$\\
2020-12-20 18:59&0.4&303.9&251.9&263.3$\pm$2.6&275.8$\pm4.1$ & 37.3& 98.7$\pm 11.6$\\
2020-12-21 05:45&0.4&307.1&251.7&266.4$\pm$2.2&274.0$\pm2.3$ & 46.8& 76.9$\pm 4.4$\\
2020-12-21 16:30&0.4&310.6&251.6&267.2$\pm$2.3&274.6$\pm4.8$ & 45.6& 79.2$\pm 9.4$\\
2020-12-22 03:14&0.4&314.2&251.4&267.1$\pm$2.1&281.4$\pm3.5$ & 28.2& 129.0$\pm 15.7$\\
2020-12-22 14:15&0.4&318.0&251.2&267.5$\pm$1.1&289.3$\pm4.7$ & 18.8& 187.2$\pm 43.7$\\
2020-12-23 00:59&0.4&321.9&251.1&266.3$\pm$1.9&283.3$\pm4.0$ & 28.4& 126.5$\pm 16.9$\\
2020-12-23 11:45&0.4&325.8&250.9&268.9$\pm$3.0&281.0$\pm4.5$ & 34.0& 106.0$\pm 13.5$\\
2020-12-23 22:30&0.4&329.7&250.8&266.6$\pm$2.9&282.5$\pm.61$ & 32.9& 109.0$\pm 1.9$\\
2020-12-24 09:14&0.4&333.6&250.6&267.1$\pm$1.5&286.5$\pm5.6$ & 28.7& 123.1$\pm 22.4$\\
2020-12-24 20:15&0.4&337.5&250.5&268.3$\pm$3.4&299.0$\pm2.8$ & 18.4& 182.5$\pm 25.6$\\
2020-12-25 06:59&0.4&341.1&250.3&267.9$\pm$2.7&293.0$\pm6.3$ & 24.7& 138.8$\pm 32.0$\\
2020-12-25 17:45&0.4&344.6&250.2&268.4$\pm$.98&296.1$\pm7.9$ & 23.6& 143.8$\pm 43.6$\\
2020-12-26 04:30&0.4&348.0&250.0&267.8$\pm$2.8&292.8$\pm5.8$ & 27.7& 123.4$\pm 23.5$\\
2020-12-26 15:14&0.4&351.1&249.9&268.5$\pm$2.6&322.2$\pm5.2$ & 11.8& 262.1$\pm 105.4$\\
2020-12-27 02:15&0.4&354.1&249.8&269.6$\pm$1.6&315.3$\pm3.3$ & 16.3& 194.7$\pm 35.4$\\
2020-12-27 12:59&0.4&356.8&249.6&269.5$\pm$3.0&323.9$\pm3.1$ & 13.8& 223.8$\pm 45.4$\\
2020-12-27 23:45&0.4&359.3&249.5&269.3$\pm$3.7&331.7$\pm6.9$ & 11.8& 254.5$\pm 135.3$\\
2020-12-28 10:30&0.4&361.6&249.4&269.7$\pm$3.7&337.8$\pm2.2$ & 10.6& 278.0$\pm 52.9$\\
2020-12-28 21:14&0.4&363.7&249.2&272.1$\pm$5.1&335.6$\pm5.0$ & 12.6& 234.1$\pm 84.0$\\
2020-12-29 08:15&0.4&365.7&249.1&271.5$\pm$3.9&343.7$\pm1.7$ & 10.5& 271.8$\pm 41.7$\\
2020-12-29 18:59&0.4&367.5&249.0&272.3$\pm$1.5&343.0$\pm1.8$ & 11.9& 240.5$\pm 33.2$\\
\enddata
\tablenotetext{a}{Number of days used to make median images}
\tablenotetext{b}{Position angle of the  anti-Sun direction, degree.}
\tablenotetext{c}{Position angle of the   negative velocity, degree.}
\tablenotetext{d}{Position angle of the dust tail, degree.}
\tablenotetext{e}{Position angle of the plasma tail, degree.}
\tablenotetext{f}{Aberration angle. }
\tablenotetext{g}{Inferred solar wind radial velocity with measured plasma tail position angles from the equation (\ref{eq:swv}). Unit:  km s$^{-1}$}
\end{deluxetable}

\clearpage

\clearpage

\clearpage

\clearpage

\clearpage

\clearpage

\clearpage

\begin{thebibliography}{}

\bibitem[Alfven(1957)]{1957Tell....9...92A} Alfven, H.\ 1957, Tellus, 9, 92. doi:10.3402/tellusa.v9i1.9064
\bibitem[Biermann(1951)]{1951ZA.....29..274B} Biermann, L.\ 1951, \zap, 29, 274
\bibitem[Biermann et al.(1967)]{1967SoPh....1..254B} Biermann, L., Brosowski, B., \& Schmidt, H.~U.\ 1967, \solphys, 1, 254. doi:10.1007/BF00150860

\bibitem[Bohren \& Huffman(1983)]{Bohren83} Bohren, C.~F. \& Huffman, D.~R.\ 1983, New York: Wiley, 1983
\bibitem[Belton \& Brandt(1966)]{1966ApJS...13..125B} Belton, M.~J.~S. \& Brandt, J.~C.\ 1966, \apjs, 13, 125. doi:10.1086/190138
\bibitem[Bothmer \& Schwenn(1998)]{1998AnGeo..16....1B} Bothmer, V. \& Schwenn, R.\ 1998, Annales Geophysicae, 16, 1. doi:10.1007/s00585-997-0001-x
\bibitem[Brandt(1968)]{1968ARA&A...6..267B} Brandt, J.~C.\ 1968, \araa, 6, 267. doi:10.1146/annurev.aa.06.090168.001411
\bibitem[Bryans \& Pesnell(2012)]{2012ApJ...760...18B} Bryans, P. \& Pesnell, W.~D.\ 2012, \apj, 760, 18. doi:10.1088/0004-637X/760/1/18
\bibitem[Brueckner et al.(1995)]{1995SoPh..162..357B} Brueckner, G.~E., Howard, R.~A., Koomen, M.~J., et al.\ 1995, \solphys, 162, 357. doi:10.1007/BF00733434
\bibitem[Buffington et al.(2008)]{2008ApJ...677..798B} Buffington, A., Bisi, M.~M., Clover, J.~M., et al.\ 2008, \apj, 677, 798. doi:10.1086/529039
\bibitem[Burlaga et al.(1990)]{1990GMS....58..373B} Burlaga, L.~F., Lepping, R.~P., \& Jones, J.~A.\ 1990, Washington DC American Geophysical Union Geophysical Monograph Series, 58, 373. doi:10.1029/GM058p0373
\bibitem[Buti(1982)]{1982ApJ...252L..43B} Buti, B.\ 1982, \apjl, 252, L43. doi:10.1086/183716
\bibitem[Buti(1982)]{1982Ap&SS..85...35B} Buti, B.\ 1982, \apss, 85, 35. doi:10.1007/BF00653430
\bibitem[Cheng et al.(2020)]{2020ApJ...897...87C} Cheng, L., Zhang, Q., Wang, Y., et al.\ 2020, \apj, 897, 87. doi:10.3847/1538-4357/ab93b6
\bibitem[Cheng et al.(2022)]{2022ApJ...928..121C} Cheng, L., Wang, Y., \& Li, X.\ 2022, \apj, 928, 121. doi:10.3847/1538-4357/ac5410

\bibitem[Clover et al.(2010)]{2010ApJ...713..394C} Clover, J.~M., Jackson, B.~V., Buffington, A., et al.\ 2010, \apj, 713, 394. doi:10.1088/0004-637X/713/1/394

\bibitem[Combi et al.(2023)]{combi23} Combi, M.~R., M{\"a}kinen, T., Bertaux, J.-L., et al.\ 2023, arXiv:2304.03333. doi:10.48550/arXiv.2304.03333
\bibitem[Cremonese et al.(1997)]{1997ApJ...490L.199C} Cremonese, G., Boehnhardt, H., Crovisier, J., et al.\ 1997, \apjl, 490, L199. doi:10.1086/311040
\bibitem[Cremonese et al.(2002)]{2002AdSpR..29.1187C} Cremonese, G., Huebner, W.~F., Rauer, H., et al.\ 2002, Advances in Space Research, 29, 1187. doi:10.1016/S0273-1177(02)00136-9

\bibitem[D{\'e}moulin \& Dasso(2009)]{2009A&A...498..551D} D{\'e}moulin, P. \& Dasso, S.\ 2009, \aap, 498, 551. doi:10.1051/0004-6361/200810971

\bibitem[Diego et al.(2020)]{2020JGRA..12528281D} Diego, P., Piersanti, M., Laurenza, M., et al.\ 2020, Journal of Geophysical Research (Space Physics), 125, e28281. doi:10.1029/2020JA028281

\bibitem[Domingo et al.(1995)]{1995SoPh..162....1D} Domingo, V., Fleck, B., \& Poland, A.~I.\ 1995, \solphys, 162, 1. doi:10.1007/BF00733425

\bibitem[Downs et al.(2013)]{2013Sci...340.1196D} Downs, C., Linker, J.~A., Miki{\'c}, Z., et al.\ 2013, Science, 340, 1196. doi:10.1126/science.1236550
\bibitem[Eselevich et al.(2001)]{2001SoPh..200..259E} Eselevich, V.~G., Fainshtein, V.~G., \& Eselevich, M.~V.\ 2001, \solphys, 200, 259. doi:10.1023/A:1010389023874

\bibitem[Eyles et al.(2009)]{2009SoPh..254..387E} Eyles, C.~J., Harrison, R.~A., Davis, C.~J., et al.\ 2009, \solphys, 254, 387. doi:10.1007/s11207-008-9299-0
\bibitem[Finson \& Probstein(1968)]{Finson68} Finson, M.~J. \& Probstein, R.~F.\ 1968, \apj, 154, 327. doi:10.1086/149761
\bibitem[Freeland \& Handy(1998)]{1998SoPh..182..497F} Freeland, S.~L. \& Handy, B.~N.\ 1998, \solphys, 182, 497. doi:10.1023/A:1005038224881
\bibitem[Froment et al.(2021)]{2021A&A...650A...5F} Froment, C., Krasnoselskikh, V., Dudok de Wit, T., et al.\ 2021, \aap, 650, A5. doi:10.1051/0004-6361/202039806
\bibitem[Gombosi et al.(1996)]{1996JGR...10115233G} Gombosi, T.~I., De Zeeuw, D.~L., H{\"a}berli, R.~M., et al.\ 1996, \jgr, 101, 15233. doi:10.1029/96JA01075
\bibitem[Gosling et al.(1987)]{1987JGR....9212399G} Gosling, J.~T., Thomsen, M.~F., Bame, S.~J., et al.\ 1987, \jgr, 92, 12399. doi:10.1029/JA092iA11p12399

\bibitem[Gosling(1990)]{1990GMS....58..343G} Gosling, J.~T.\ 1990, Washington DC American Geophysical Union Geophysical Monograph Series, 58, 343. doi:10.1029/GM058p0343
\bibitem[Gosling \& Pizzo(1999)]{1999SSRv...89...21G} Gosling, J.~T. \& Pizzo, V.~J.\ 1999, \ssr, 89, 21. doi:10.1023/A:1005291711900
\bibitem[Goetz et al.(2019)]{2019A&A...630A..38G} Goetz, C., Tsurutani, B.~T., Henri, P., et al.\ 2019, \aap, 630, A38. doi:10.1051/0004-6361/201833544

\bibitem[G{\"o}tz et al.(2022)]{2022arXiv221104887G} G{\"o}tz, C., Deca, J., Mandt, K., et al.\ 2022, arXiv:2211.04887
\bibitem[Hoeksema et al.(1982)]{1982JGR....8710331H} Hoeksema, J.~T., Wilcox, J.~M., \& Scherrer, P.~H.\ 1982, \jgr, 87, 10331. doi:10.1029/JA087iA12p10331

\bibitem[Howard et al.(2008)]{2008SSRv..136...67H} Howard, R.~A., Moses, J.~D., Vourlidas, A., et al.\ 2008, \ssr, 136, 67. doi:10.1007/s11214-008-9341-4
\bibitem[Hui(2023)]{2023AJ....165...94H} Hui, M.-T.\ 2023, \aj, 165, 94. doi:10.3847/1538-3881/acae9c

\bibitem[Ip \& Mendis(1976)]{1976Icar...29..147I} Ip, W.-H. \& Mendis, D.~A.\ 1976, \icarus, 29, 147. doi:10.1016/0019-1035(76)90110-X
\bibitem[Kaiser et al.(2008)]{2008SSRv..136....5K} Kaiser, M.~L., Kucera, T.~A., Davila, J.~M., et al.\ 2008, \ssr, 136, 5. doi:10.1007/s11214-007-9277-0

\bibitem[Khabarova \& Obridko(2012)]{2012ApJ...761...82K} Khabarova, O. \& Obridko, V.\ 2012, \apj, 761, 82. doi:10.1088/0004-637X/761/2/82
\bibitem[Kilpua et al.(2017)]{2017LRSP...14....5K} Kilpua, E., Koskinen, H.~E.~J., \& Pulkkinen, T.~I.\ 2017, Living Reviews in Solar Physics, 14, 5. doi:10.1007/s41116-017-0009-6

\bibitem[Liu et al.(2005)]{2005P&SS...53....3L} Liu, Y., Richardson, J.~D., \& Belcher, J.~W.\ 2005, \planss, 53, 3. doi:10.1016/j.pss.2004.09.023

\bibitem[Horanyi(1996)]{1996ARA&A..34..383H} Horanyi, M.\ 1996, \araa, 34, 383. doi:10.1146/annurev.astro.34.1.383
\bibitem[Howard et al.(2008)]{2008SSRv..136...67H} Howard, R.~A., Moses, J.~D., Vourlidas, A., et al.\ 2008, \ssr, 136, 67. doi:10.1007/s11214-008-9341-4
\bibitem[Hundhausen(1968)]{1968SSRv....8..690H} Hundhausen, A.~J.\ 1968, \ssr, 8, 690. doi:10.1007/BF00175116
\bibitem[Jackson et al.(2013)]{2013AIPC.1539..364J} Jackson, B.~V., Buffington, A., Clover, J.~M., et al.\ 2013, Solar Wind 13, 1539, 364. doi:10.1063/1.4811062
\bibitem[Jewitt \& Li(2010)]{2010AJ....140.1519J} Jewitt, D. \& Li, J.\ 2010, \aj, 140, 1519. doi:10.1088/0004-6256/140/5/1519
\bibitem[Jia et al.(2009)]{2009ApJ...696L..56J} Jia, Y.~D., Russell, C.~T., Jian, L.~K., et al.\ 2009, \apjl, 696, L56. doi:10.1088/0004-637X/696/1/L56
\bibitem[Jones et al.(2018)]{2018SSRv..214...20J} Jones, G.~H., Knight, M.~M., Battams, K., et al.\ 2018, \ssr, 214, 20. doi:10.1007/s11214-017-0446-5
\bibitem[Kaiser et al.(2008)]{2008SSRv..136....5K} Kaiser, M.~L., Kucera, T.~A., Davila, J.~M., et al.\ 2008, \ssr, 136, 5. doi:10.1007/s11214-007-9277-0
\bibitem[Kilpua et al.(2017)]{2017LRSP...14....5K} Kilpua, E., Koskinen, H.~E.~J., \& Pulkkinen, T.~I.\ 2017, Living Reviews in Solar Physics, 14, 5. doi:10.1007/s41116-017-0009-6
\bibitem[Kohl et al.(1995)]{1995SoPh..162..313K} Kohl, J.~L., Esser, R., Gardner, L.~D., et al.\ 1995, \solphys, 162, 313. doi:10.1007/BF00733433

\bibitem[Ledoux et al.(2001)]{Ledoux01} Ledoux, G., Guillois, O., Huisken, F., et al.\ 2001, \aap, 377, 707. doi:10.1051/0004-6361:20011136

\bibitem[Liu et al.(2005)]{2005P&SS...53....3L} Liu, Y., Richardson, J.~D., \& Belcher, J.~W.\ 2005, \planss, 53, 3. doi:10.1016/j.pss.2004.09.023

\bibitem[Niedner \& Brandt(1978)]{1978ApJ...223..655N} Niedner, M.~B. \& Brandt, J.~C.\ 1978, \apj, 223, 655. doi:10.1086/156299
\bibitem[Niedner \& Brandt(1979)]{1979ApJ...234..723N} Niedner, M.~B. \& Brandt, J.~C.\ 1979, \apj, 234, 723. doi:10.1086/157549
\bibitem[Niedner et al.(1978)]{1978ApJ...221.1014N} Niedner, M.~B., Rothe, E.~D., \& Brandt, J.~C.\ 1978, \apj, 221, 1014. doi:10.1086/156107

\bibitem[Mann et al.(2007)]{Mann07} Mann, I., Murad, E., \& Czechowski, A.\ 2007, \planss, 55, 1000. doi:10.1016/j.pss.2006.11.015
\bibitem[Maunder et al.(2022)]{2022SoPh..297..148M} Maunder, M.~L., Foullon, C., Forsyth, R., et al.\ 2022, \solphys, 297, 148. doi:10.1007/s11207-022-02077-3

\bibitem[Meng et al.(2022)]{2022RAA....22c5018M} Meng, M.-M., Liu, Y.~D., Chen, C., et al.\ 2022, Research in Astronomy and Astrophysics, 22, 035018. doi:10.1088/1674-4527/ac49e4
\bibitem[Morrill et al.(2006)]{2006SoPh..233..331M} Morrill, J.~S., Korendyke, C.~M., Brueckner, G.~E., et al.\ 2006, \solphys, 233, 331. doi:10.1007/s11207-006-2058-1
\bibitem[Nistic{\`o} et al.(2022)]{2022ApJ...938...20N} Nistic{\`o}, G., Zimbardo, G., Perri, S., et al.\ 2022, \apj, 938, 20. doi:10.3847/1538-4357/ac8e62

\bibitem[Notni \& Tiersch(1987)]{1987A&A...187..796N} Notni, P. \& Tiersch, H.\ 1987, \aap, 187, 796
\bibitem[Omidi \& Winske(1987)]{1987JGR....9213409O} Omidi, N. \& Winske, D.\ 1987, \jgr, 92, 13409. doi:10.1029/JA092iA12p13409

\bibitem[Owens \& Cargill(2004)]{2004AnGeo..22.4397O} Owens, M. \& Cargill, P.\ 2004, Annales Geophysicae, 22, 4397. doi:10.5194/angeo-22-4397-2004

\bibitem[Oyama \& Hirao(1985)]{1985AdSpR...5l..65O} Oyama, K.~I. \& Hirao, K.\ 1985, Advances in Space Research, 5, 65. doi:10.1016/0273-1177(85)90068-7
\bibitem[Parker(1958)]{1958ApJ...128..664P} Parker, E.~N.\ 1958, \apj, 128, 664. doi:10.1086/146579
\bibitem[Pecora et al.(2022)]{2022ApJ...929L..10P} Pecora, F., Matthaeus, W.~H., Primavera, L., et al.\ 2022, \apjl, 929, L10. doi:10.3847/2041-8213/ac62d4
\bibitem[Poirier et al.(2023)]{2023arXiv230705294P} Poirier, N., R{\'e}ville, V., Rouillard, A.~P., et al.\ 2023, arXiv:2307.05294. doi:10.48550/arXiv.2307.05294

\bibitem[Povich et al.(2003)]{2003Sci...302.1949P} Povich, M.~S., Raymond, J.~C., Jones, G.~H., et al.\ 2003, Science, 302, 1949. doi:10.1126/science.1092142
\bibitem[Ragot \& Kahler(2003)]{2003ApJ...594.1049R} Ragot, B.~R. \& Kahler, S.~W.\ 2003, \apj, 594, 1049. doi:10.1086/377076
\bibitem[Ramanjooloo \& Jones(2022)]{2022JGRA..12729799R} Ramanjooloo, Y. \& Jones, G.~H.\ 2022, Journal of Geophysical Research (Space Physics), 127, e29799. doi:10.1029/2021JA029799
\bibitem[Raymond et al.(1998)]{1998ApJ...508..410R} Raymond, J.~C., Fineschi, S., Smith, P.~L., et al.\ 1998, \apj, 508, 410. doi:10.1086/306391
\bibitem[Raymond et al.(2018)]{2018ApJ...858...19R} Raymond, J.~C., Downs, C., Knight, M.~M., et al.\ 2018, \apj, 858, 19. doi:10.3847/1538-4357/aabade
\bibitem[Raymond et al.(2022)]{2022ApJ...926...93R} Raymond, J.~C., Giordano, S., Mancuso, S., et al.\ 2022, \apj, 926, 93. doi:10.3847/1538-4357/ac3cbd
\bibitem[Richardson \& Cane(2010)]{2010SoPh..264..189R} Richardson, I.~G. \& Cane, H.~V.\ 2010, \solphys, 264, 189. doi:10.1007/s11207-010-9568-6
\bibitem[Tasnim \& Cairns(2016)]{2016JGRA..121.4966T} Tasnim, S. \& Cairns, I.~H.\ 2016, Journal of Geophysical Research (Space Physics), 121, 4966. doi:10.1002/2016JA022725

\bibitem[Thomas(2020)]{2020icpr.book.....T} Thomas, N.\ 2020, An Introduction to Comets; Post-Rosetta Perspectives. ISBN: 978-3-030-50574-5. Cham: Springer International Publishing, 2020.. doi:10.1007/978-3-030-50574-5, Chapter 5
\bibitem[Thompson \& Wei(2010)]{2010SoPh..261..215T} Thompson, W.~T. \& Wei, K.\ 2010, \solphys, 261, 215. doi:10.1007/s11207-009-9476-9

\bibitem[Utterback \& Kissel(1990)]{Utterback90} Utterback, N.~G. \& Kissel, J.\ 1990, \aj, 100, 1315. doi:10.1086/115599


\bibitem[Uzzo et al.(2001)]{2001ApJ...558..403U} Uzzo, M., Raymond, J.~C., Biesecker, D., et al.\ 2001, \apj, 558, 403. doi:10.1086/322473
\bibitem[Valenzuela et al.(1986)]{1986Natur.320..700V} Valenzuela, A., Haerendel, G., F{\"o}ppl, H., et al.\ 1986, \nat, 320, 700. doi:10.1038/320700a0
\bibitem[Voelzke(2005)]{2005EM&P...97..399V} Voelzke, M.~R.\ 2005, Earth Moon and Planets, 97, 399. doi:10.1007/s11038-006-9073-y
\bibitem[Vourlidas et al.(2007)]{2007ApJ...668L..79V} Vourlidas, A., Davis, C.~J., Eyles, C.~J., et al.\ 2007, \apjl, 668, L79. doi:10.1086/522587
\bibitem[Weber \& Davis(1967)]{1967ApJ...148..217W} Weber, E.~J. \& Davis, L.\ 1967, \apj, 148, 217. doi:10.1086/149138
\bibitem[Wegmann(2000)]{2000A&A...358..759W} Wegmann, R.\ 2000, \aap, 358, 759
\bibitem[Witasse et al.(2017)]{2017JGRA..122.7865W} Witasse, O., S{\'a}nchez-Cano, B., Mays, M.~L., et al.\ 2017, Journal of Geophysical Research (Space Physics), 122, 7865. doi:10.1002/2017JA023884

\bibitem[Witt et al.(1998)]{Witt98} Witt, A.~N., Gordon, K.~D., \& Furton, D.~G.\ 1998, \apjl, 501, L111. doi:10.1086/311453
\bibitem[Wu \& Lepping(2011)]{2011SoPh..269..141W} Wu, C.-C. \& Lepping, R.~P.\ 2011, \solphys, 269, 141. doi:10.1007/s11207-010-9684-3
\bibitem[Zhang et al.(2023)]{2023PSJ.....4...70Z} Zhang, Q., Battams, K., Ye, Q., et al.\ 2023, The Planetary Science Journal, 4, 70. doi:10.3847/PSJ/acc866

\end{thebibliography}
\end{document}